\documentclass[prb,reprint,superscriptaddress,
amsmath,amssymb,aps,]{revtex4-2}
\usepackage{array}
\usepackage{amsmath}
\usepackage{graphicx}
\usepackage{dcolumn}
\usepackage{bm}
\usepackage{float}
\usepackage{hyperref}
\usepackage{physics}
\usepackage{tikz}
\hypersetup{colorlinks,linkcolor={blue},citecolor={blue},urlcolor={blue}}  
\usepackage{xtab,afterpage,longtable}
\usepackage[utf8]{inputenc}
\usepackage{comment}

\DeclareUnicodeCharacter{03C3}{\ensuremath{\sigma}}
\makeatletter
\def\LT@LR@e{\LTleft\z@   \LTright\z@}%
\makeatother

\begin{document}

\preprint{APS/123-QED}

\title{Green's Functions from Sample-based Krylov Quantum Diagonalization: An Impurity Solver for Dynamical Mean-Field Theory}
\author{Jay Patel}
\affiliation{Center for Computation and Technology, Louisiana State University, Baton Rouge, LA 70803, USA}
\affiliation{Department of Physics, The Pennsylvania State University, University Park, PA 16802, USA}
\author{Chakradhar Rangi}
\affiliation{Department of Physics and Astronomy, Louisiana State University, Baton Rouge, LA 70803, USA}
\author{Ka-Ming Tam}
\email{kmtam@lsu.edu}
\affiliation{Center for Computation and Technology, Louisiana State University, Baton Rouge, LA 70803, USA}
\affiliation{Department of Physics and Astronomy, Louisiana State University, Baton Rouge, LA 70803, USA}
\date{\today}

\begin{abstract}

We generalize the sample-based Krylov quantum diagonalization (SKQD) method from ground-state calculations to the evaluation of single-particle Green’s functions. By constructing and sampling unitary Krylov subspaces in the $N\pm 1$ particle-number sectors and evaluating all sector-connecting overlaps classically, the approach reconstructs the Green’s function via a Lanczos continued fraction while retaining the shallow-circuit, ancilla-free character of SKQD. The quantum device is required only to prepare and sample short-time evolutions. Applied to the particle-hole-symmetric single-impurity Anderson model in chain geometry, with the discrete bath representation used in dynamical mean-field theory (DMFT), the method recovers the spectral function using a relatively small fraction of the full Hilbert space. Across a range of interaction strengths that spans the metal–insulator transition, the main spectral features are reproduced. These results suggest that SKQD-based Green’s-function calculations may allow DMFT impurity solvers with a larger number of bath sites on near-term quantum hardware than is currently practical.
\end{abstract}\maketitle

\section{Introduction}

Strongly correlated materials such as transition metal oxides, heavy fermion compounds, and cuprates are defined by a Coulomb repulsion comparable to the kinetic energy scale, so that neither limit is a perturbation of the other and single-particle methods fail qualitatively \cite{Georges1996,Kotliar2006}. Exact treatment is obstructed by the exponential growth of the Hilbert-space dimension, and thus exact diagonalization (ED) is limited to modest system sizes.

The ground state is the reference from which excitations are defined, yet it is not itself what experiments usually measure. Angle-resolved photoemission, scanning tunneling spectroscopy, and optical conductivity all probe the single-particle Green's function and its spectral density. The quantities that organize our understanding of correlated matter, such as the quasiparticle residue, the Hubbard bands, and the Mott gap, are spectral rather than ground state properties. A method that aspires to make contact with experiment must therefore deliver excitation spectra or frequency-dependent Green's functions.
Classically, Green's functions of correlated models are obtained from ED, quantum Monte Carlo (QMC), numerical renormalization group, or, more recently, tensor-network solvers \cite{Gull2011,Bulla2008}, each with a hard limitation. ED is confined to small clusters; QMC faces a fermionic sign problem away from special, typically particle-hole symmetric points and returns imaginary-axis data requiring ill-conditioned analytic continuation; tensor networks are constrained by entanglement growth beyond one dimension and at long times. Within dynamical mean-field theory (DMFT) \cite{Metzner1989,Georges1996}, which maps the lattice onto a single interacting impurity in a self-consistently determined bath, these limitations reappear in the so-called impurity solver, the only many-body step of the DMFT loop. Hamiltonian-based solvers, such as ED, yield the full real-frequency spectral function but afford only a handful of bath orbitals \cite{Caffarel1994}, so that bath discretization, rather than the physics, often sets the resolution. The situation is further strained by disorder, as in the Anderson–Hubbard model, the interplay of randomness and interactions often defeats simple perturbative solvers, and capturing localization within a local theory requires typical-medium constructions with fast and accurate impurity solvers \cite{Dobrosavljevic2003,Terletska2018,Zhang2015,Ekuma2015,Tam2021}. An impurity solver delivering accurate Green's functions for a substantially larger number of bath sites would directly enlarge the reach of DMFT and its cluster, nonequilibrium, and typical-medium extensions \cite{Aoki2014,Dohner2022,Rangi2025,Rangi2024}.

Quantum computers provide a new avenue for studying correlated systems that bypasses some of these limitations. The variational quantum eigensolver can be performed in a shallow circuit but requires nonconvex optimization prone to barren plateaus \cite{Peruzzo2014,McClean2018}. Krylov methods are viable alternatives. They require no explicit optimization and are built from shallow real-time evolution circuits \cite{Stair2020,Klymko2022,Cortes2022,Yoshioka2025}. The recently proposed Sample-based Krylov quantum diagonalization (SKQD) \cite{Klymko2022,Yu2025,Piccinelli2025} improves its applicability in present quantum hardware. We note  that other diagonalization methods in restricted Hilbert space have also been proposed over the past few years \cite{Kanno2026,Mikkelsen2025,Sugisaki2025,Motta2024,Reinholdt2025,RobledoMoreno2025}.

Calculations of the Green's functions are more challenging as their poles span the excitation spectra of two particle-number sectors, and their residues require operator overlaps connecting those sectors. Existing quantum algorithms for Green's functions reconstruct them from real-time correlators measured via an ancilla qubit for Hadamard tests, from variational compilation of the propagator, from quantum equation-of-motion or subspace-expansion techniques, or from a Lanczos recursion driven by measured moments \cite{Endo2020,Rizzo2022,Kanasugi2023,Baker2021,Rungger2019,Jamet2021,Jamet2022,GreeneDiniz2024,Keen2021,Bishop2023}. A complementary ancilla-free approach based on a linear-response framework has also been demonstrated on quantum hardware \cite{Kokcu2024}. It remains unclear whether the sampling-based Krylov construction can be extended from eigenenergies to the full Lehmann representation, and whether the extreme basis compression that suffices for ground states survives for spectral functions.

In this work we generalize SKQD to the single-particle Green's functions by diagonalizing the Hamiltonian in sampled Krylov spaces of the $N{\pm}1$-particle sectors and assembling the Green's functions from the resulting poles and residues, with all sector-connecting overlaps evaluated classically from the sampled configurations. The quantum device is only needed to prepare and sample short-time evolutions, so the construction retains the shallow-circuit, ancilla-free character of SKQD. Applied to the particle-hole-symmetric Anderson impurity in chain geometry as the natural representation of the DMFT bath, the method reproduces the spectral function with a basis comprising only a small fraction of the Hilbert space; it captures the character of the interaction-driven metal-insulator transition as $U$ increases. 

The remainder of the paper is organized as follows. Section \ref{sec:Model} introduces the single-impurity Anderson model. Section \ref{sec:Method} reviews the path from classical power-Krylov spaces to unitary quantum Krylov evolution and SKQD, the choice of time step and Krylov dimension, and then presents our algorithm for Green’s functions. Section \ref{sec:Results} presents the spectral function and its convergence as a function of the number of sampled basis states. Section \ref{sec:Conclusion} concludes with an outlook for embedding the solver in a full DMFT loop on quantum hardware. Three additional sections are provided in the appendix for the detail of the Lanczos algorithm for generating the Green's functions in the truncated Hilbert space, additional parameters used, and data generated for different numbers of bath sites.


\section{Model}
\label{sec:Model}

We focus on the single-impurity Anderson model (SIAM), for which a discrete bath approximation is used so that it can be represented in a finite number of so-called bath sites. We define the model in the chain geometry: an interacting impurity at site
$0$ attached to a 1D chain of non-interacting bath sites with nearest-neighbor hopping. We assume there is a single impurity at site 0, with $L-1$ bath sites to represent the electron bath. Hence there are $2L$ spin-orbitals, which require $2L$ qubits. The Hamiltonian can be written as
\begin{eqnarray}
H = & U n_{0\uparrow}n_{0\downarrow} + \epsilon_{d}\sum_{\sigma}n_{0\sigma} 
    + \sum_{i=1}^{L-1}\sum_{\sigma}\epsilon_{i}n_{i\sigma} 
        \\ \nonumber
    - &\sum_{i=0}^{L-2}\sum_{\sigma}t_{i}\!
      \left(c_{i\sigma}^{\dagger} c_{i+1\sigma}+\text{H.c.}\right),
  \label{eq:siam}
\end{eqnarray}
where $c_{i\sigma}$ and $c_{i\sigma}^{\dagger}$ are the annihilation and creation operators of an electron at site $i$ with spin $\sigma = \uparrow / \downarrow$, $n_{i\sigma}$ is the density operator for site $i$ with spin $\sigma$. $U$ is the interaction, $\epsilon_d$ is the local energy of the impurity, $\epsilon_i$ is the bath site local energy for the $i$-th bath site, $t_i$ is the hopping matrix element between the $i$-th and $(i+1)$-th sites. We denote the number of up and down electrons as $N_{\uparrow}$ and $N_{\downarrow}$ respectively and $N=N_{\uparrow}+N_{\downarrow}$ as the total number of electrons. We only consider the half-filled, particle-hole symmetric, spin-balanced case so that we fix $N_{\uparrow} = N_{\downarrow} = L/2$. All energies are quoted in units of the hopping $t = 1$ used in the DMFT bath fit.

For this half-filled, particle-hole symmetric case, there is no local potential for the bath sites arranged in the chain geometry; in general the local potential can be non-zero. For convenience in discussing the transformation to different bases, we can rewrite the Hamiltonian as  
\begin{equation}H = \sum_{ij\sigma} h_{ij}\, c^\dagger_{i\sigma}c_{j\sigma} + U\, n_{0\uparrow}n_{0\downarrow},\end{equation} with 
\begin{equation}\qquad h_{00}=\epsilon_d,\ h_{i\space i+1}=h_{i+1 \space i}=-t_i.\end{equation}


\section{Method}
\label{sec:Method}
Naively, SKQD works only if the weight of the ground-state wave function is concentrated on a small fraction of the basis states in a large Hilbert space. If that weight is distributed nearly uniformly, SKQD does not offer a substantial computational advantage. The underlying Krylov method would still work; however, extracting the Krylov space by sampling would not help, because the number of samples required would not be substantially smaller than the Hilbert space dimension. A simple choice is to start from the bare basis in which the quadratic part of the Hamiltonian is diagonalized. For weak interaction one might expect the Hartree–Fock ground state to be a better starting point. We test both the bare basis and the Hartree–Fock (HF) basis and find that the HF basis is not generally superior for SKQD as far as the spectral function is concerned. Because the choice of basis affects how the impurity Green’s function is computed, we first recap the ground-state algorithm from the literature \cite{Klymko2022,Yu2025,Piccinelli2025}. This serves as a brief review of SKQD and fixes the notation for the Green’s-function calculation.

We first solve the $L\times L$ non-interacting bare problem or Hartree–Fock problem. At particle–hole symmetric point, self-consistency gives $\langle n_{0\sigma}\rangle = 1/2$. Therefore $h_{00}=h^{\rm HF}_{00} = \epsilon_d + U/2 = 0$. For the non-interacting bare case, $h_{00}=h^{\rm Bare}_{00} = \epsilon_d = -U/2$. Note that the latter does not obey particle-hole symmetry. We diagonalize the quadratic terms of the Hamiltonian \begin{equation}W^{-1}h W = \,\mathrm{diag}(\varepsilon),\end{equation} where the columns of $W$ are the eigenvectors and $\varepsilon$ are the eigenvalues of $h$. We can then define the new basis as \begin{equation}\qquad d^\dagger_{p\sigma} = \sum_i W_{ip}c^\dagger_{i\sigma}.\end{equation}

 The Hamiltonian in the new basis becomes \begin{equation}H = \sum_{pq\sigma}\tilde h_{pq}\,d^\dagger_{p\sigma}d_{q\sigma} + U\hat N_\uparrow\hat N_\downarrow,\end{equation}
where $\tilde h$ is given as \begin{equation}\tilde h = W^{-1}hW.\end{equation} The impurity density in the bare basis is  \begin{equation}\hat N_\sigma = \sum_{pq}P_{pq}d^\dagger_{p\sigma}d_{q\sigma},\end{equation}with \begin{equation}\ P=\mathbf{w}\mathbf{w}^T,\end{equation}
where $\mathbf{w} \equiv W_{0,:}.$ We can also write the inverse transform, for example the creation operator for the impurity in the bare basis as
\begin{equation}c^\dagger_{0\uparrow} = \sum_p w_p\, d^\dagger_{p\uparrow}.\end{equation}This also indicates that we need to consider multiple-mode excitations when we calculate the spectral function in the bare basis. 

\subsection{Classical Krylov Method}
With the above discussion of the transformation to the bare basis, we then discuss the Krylov method. The main idea is to find a new basis for the full Hamiltonian in which the ground state can be represented approximately in a smaller dimension than that of the full Hilbert space. 

The standard Krylov method generates the basis by applying $H$ repeatedly to an initial state. Any state in a given Hilbert space can in principle be written in its eigencomponent decomposition, \begin{equation}\ket{\psi_{0}}=\sum_{i}\gamma_{i}\ket{\phi_{i}},\end{equation} where $|\phi_i\rangle$ are the eigenstates and the $\gamma_i$ are arbitrary real coefficients. Since
\begin{equation}H^{k}\ket{\psi_{0}}=\sum_{i}\gamma_{i}E_{i}^{k}\ket{\phi_{i}},\end{equation} the component of the extremal eigenvector is amplified geometrically by the power of eigenvalues $E_i^k$. By retaining intermediates during the iteration process we form the Krylov space with dimension $d$ defined as 
\begin{equation}
  Kr_{d}(H,\ket{\psi_{0}}) \equiv \operatorname{span}
    \{\ket{\psi_{0}},H\ket{\psi_{0}},\dots,H^{d-1}\ket{\psi_{0}}\}.
\end{equation}
One can then build the Hamiltonian in the Krylov basis
$\widetilde{H}_{jk}=\langle{\psi_{j}}|{H}|{\psi_{k}}\rangle$, as well as the overlap of the basis vectors 
$\widetilde{S}_{jk}=\braket{\psi_{j}}{\psi_{k}}$ and solve the generalized eigenvalue problem 
$\widetilde{H}v=\widetilde{E}\widetilde{S}v$. 
The method is variational and should provide an improvement over the monomial $H^{d-1}$ as that in the power method \cite{Saad2003}. 

Note that the $H^{k}\ket{\psi_{0}}$ all rotate toward the dominant eigenvector and become near-parallel in the Hilbert space, so $\widetilde{S}$ is severely ill-conditioned and requires truncation or reorthogonalization \cite{Saad2003}. 

\subsection{Quantum Krylov Method}
Classical Krylov and related Lanczos techniques have long been standard for computing single-particle Green's functions of impurity models via continued fractions or the Lehmann representation \cite{Caffarel1994,Bulla2008}. However, the method is not readily implemented on quantum hardware because the projection is not unitary. An alternative is to replace the projection by real-time evolution. The method is referred to as
Unitary Krylov Quantum Diagonalization (KQD) \cite{Stair2020,Klymko2022,Cortes2022,Yoshioka2025}.

Replacing the power of Hamiltonian $H^{k}$ by real-time evolutions, we have
\begin{equation}
  \ket{\psi_{k}} \equiv  e^{-ikH\Delta t}\ket{\psi_{0}},
  \label{eq:krylovstates}
\end{equation}
for $ \qquad k=0,\dots,K-1$. This is implementable on quantum hardware with the Trotter approximation. As it is unitary and thus norm-preserving, the basis cannot contract. The space spanned by this basis consists of trigonometric rather than polynomial filters in $H$. We refer the reader to the literature for a discussion of the convergence \cite{Stair2020,Klymko2022,Cortes2022,Epperly2022,Kirby2023,Kirby2024,Lee2024,Shen2023}. The optimal choice of $\Delta t$ clearly depends on the system being studied. Here we set
\begin{equation}
  \Delta t = \pi/(2U),
  \label{eq:dt}
\end{equation}
essentially treating $U$ as the dominant scale. Too large a $\Delta t$ aliases well-separated "eigenphases" onto each other and the subspace resolution
saturates below $d$. Too small and the states are all nearly $\ket{\psi_{0}}$ and cannot explore the Hilbert space.

The major challenge of implementing KQD is that $\widetilde{H}_{jk}$ is not an expectation value in a single state, since it connects two distinct Krylov vectors, and usually requires a Hadamard test.
The SKQD is designed to mitigate the difficulty by sampling the basis generated by the evolution \cite{Klymko2022,RobledoMoreno2025}. Instead of forming the matrix elements for the Hamiltonian and the overlap of the basis, one first samples the basis being generated itself. The idea is that the distribution of the coefficients from the basis generated by the unitary KQD method is dominated by a small fraction of the basis states compared to the full Hilbert space. This suggests that the factorizability or the related entanglement of the many-body wave function dictates the quality of SKQD.

The quantum measurement for SKQD is essentially a sampling of the weight of each basis being generated. This can be done by measuring all $2L$ qubits in the computational basis, and repeating this $M$ times for statistics. The advantage of the SKQD is that it avoids phase-estimation primitives such as the Hadamard test, which have proved challenging on NISQ devices. In a sense the method is a ‘true’ quantum Monte Carlo scheme: the stochastic nature of the computational-basis measurements yields an estimate of the state from the sampled configuration counts.

\subsection{SKQD for Green's function}
Since we are interested in the calculation of the Green's function, this requires Krylov bases for three sectors with different particle numbers. These are the $N$-particle sector, the half-filled plus one spin-up sector, and the half-filled minus one spin-up sector. For the particle-hole symmetric case, we only need either the plus one or minus one sector as the other can be inferred by using particle-hole symmetry. We also assume there is no symmetry breaking in the spin sector, so that we only need to calculate the Green's function for either the spin up or spin down sector. We denote these three basis sets as
\begin{equation}|\phi^{\mathcal G}_0\rangle = |\phi_{\rm ref}\rangle, \qquad |\phi^{\mathcal A}_0\rangle \propto c^\dagger_{0\uparrow}|\phi_{\rm ref}\rangle, \qquad |\phi^{\mathcal R}_0\rangle \propto c_{0\uparrow}|\phi_{\rm ref}\rangle,\end{equation}
where $|\phi_{ref}\rangle$ is given by either the non-interacting bare or Hartree-Fock ground state.

For each sector $S={\mathcal G, \mathcal A, \mathcal R}$ and each $k = 0,\dots,K-1$, we apply the Trotterized propagation $\prod_{k}e^{-iH\Delta t}$ to the $|\phi_{ref}\rangle$. Once the basis is formed, we can measure all $2L$ qubits in the computational basis, repeat $M$ times for statistics. 
\begin{equation}\mathcal{S}_S = \bigcup_{k=0}^{K-1}\big\{\,b^{(1)}_k,\dots,b^{(M)}_k\,\big\},\end{equation}
where \begin{equation}\qquad b^{(j)}_k \sim |\langle b|\phi^S_k\rangle|^2.\end{equation} We can understand $b$ as bitstrings composed of $2L$ classical bits.


With the unitary Krylov basis sets obtained for the ground state sector $\mathcal{S}_{\mathcal G}$, the Hamiltonian is then projected onto the sampled subspace and it reduces to solving the Rayleigh–Ritz problem as follows,
\begin{equation}H^{\mathcal S}_{bb'} = \langle b|H|b'\rangle,\end{equation} where
\begin{equation}b,b'\in\mathcal{S}_\mathcal{G}.\end{equation} Solving the eigenvalue problem, we have
\begin{equation}H^{\mathcal S}\boldsymbol v = E_0\boldsymbol v, \quad |\Psi_0\rangle=\sum_b v_b|b\rangle,\end{equation}
where $E_0$ and $\boldsymbol{v} = (v_0, v_1, v_2, \cdots)^T$ are the ground state energy and the ground state eigenvector respectively. We denote the ground state wavefunction as $|\Psi_0\rangle$. This step is done by classical hardware, as the dimension of the basis is presumably small enough to be handled by classical computation either by full diagonalization or  Lanczos method for the ground state. Note that the SKQD projected Hamiltonian is an eigenvalue problem unlike a generalized eigenvalue problem of KQD. 
 
In order to fulfill the sum rule, that is neither to gain nor to lose the spectral weight, we need to augment the response subspaces as 
\begin{equation}
\label{eq:Aspace}
\mathcal{A}^{\pm} \equiv \mathcal{S}_{\mathcal{A},\mathcal{R}} \ \cup\ \!\big(c^{(\dagger)}_{0\uparrow}\,\mathcal{S}_\mathcal{G}\big), \end{equation} and
\begin{equation}\big(c^\dagger_{0\uparrow}\mathcal{S}_\mathcal{G}\big) = \bigcup_{b\in\mathcal{S}_\mathcal{G}}\ \bigcup_{p:\,w_p\neq0}\big\{\,d^\dagger_{p\uparrow}|b\rangle\,\big\},\end{equation}
where the condition $p:\,w_p\neq0$ is to exclude those basis states with vanishing weight.
This inclusion of the basis set is to make $\chi^\pm \equiv c^{(\dagger)}_{0\uparrow}|\Psi_0\rangle$ lie entirely inside the basis $\mathcal{A}^\pm$, so that $\hat P_{\mathcal{A}}|\chi\rangle = |\chi\rangle$. With this, we can show that  \begin{equation}\sum_m |Z^+_m|^2 = \langle\chi^+|\chi^+\rangle = 1 - \langle \Psi_0| n_{0\uparrow}|{\Psi_0}\rangle,\end{equation}
where $Z^{+}_{m} = \langle m^{(N+1)}|c^{\dagger}_{0,\uparrow}|\Psi_{0}\rangle$ is the spectral weight of the m-th pole of the addition branch, with $|m^{(N+1)}\rangle$ the eigenstates of $H$ in the $N+1$ particle sector and $|\Psi_{0}\rangle$ is the ground state in the $N$ particle sector. This relation is satisfied for any pool size chosen for the SKQD approximation. It guarantees that the spectral weight is redistributed among poles but never lost. 

The inner union over $p$ is the price of the rotated basis, the new basis is a linear combination of the original basis, up to $L - N_{\uparrow/\downarrow}$ children per parent instead of one. Note that \begin{equation}|\chi^+\rangle = c^\dagger_{0\uparrow}|\Psi_0\rangle = \sum_{b\in\mathcal{S}_\mathcal{G}}v_b \sum_{p} w_p\, d^\dagger_{p\uparrow}|b\rangle\end{equation} restricted to $\mathcal{A}^+$; likewise $|\chi^-\rangle$ with $c_{0\uparrow}$ on $\mathcal{A}^-$. 

At this point we have the ingredients needed to form the Green's function. We can either use the Lehmann spectral representation formula, or we can use the Lanczos method inside $\mathcal{A}^\pm$ with seed $|\chi^\pm\rangle/\|\chi^\pm\|$ for $n_L$ steps, giving coefficients $\{\alpha_j,\beta_j\}$ for the tridiagonal matrix generated by the Lanczos method \cite{Lin1993}. The particle component of the Green's function can then be written as  

\begin{equation}
\label{eq:Lanczos_CF}
G^+(\omega) = \cfrac{\|\chi^+\|^2}{z - \alpha_0 - \cfrac{\beta_1^2}{z - \alpha_1 - \cfrac{\beta_2^2}{\ddots}}}\ ,\end{equation}
where $\qquad z = \omega + E_0 + i\eta$. And similarly for the hole component $G^-(\omega)$ by replacing  $\|\chi^+\|^2$ with  $\|\chi^-\|^2$ and $z = E_0 +\omega + i\eta$ by $z' = E_0 - \omega - i\eta$ as well as an overall minus sign. The spectral function can be obtained via the retarded Green's function $G^{ret} = G^+ + G^-$ as
\begin{equation}A(\omega) = -\tfrac{1}{\pi}\operatorname{Im}\big[G^+(\omega)+G^-(\omega)\big].\end{equation}

We provide a short discussion of the Lanczos method applied to this truncated space from sampling, in particular the recursive relations for finding $\{\alpha_j,\beta_j\}$ in appendix \ref{app:lanczos}. The computational cost scales as $\mathcal O(d^2 n_L)$ rather than $\mathcal O(d^3)$ for the Lehmann representation. The Lanczos expansion order $n_L$ is typically of order $\mathcal O(10^2)$, and $d$ can be a rather large number compared to $n_L$, particularly if the number of bath sites is large, which is the purpose of the present algorithm. 

\subsection{Implementation and use of AI tools}
The numerical implementation and the plotting scripts were developed with the assistance of large language models (Anthropic Claude Opus 4.8 and
Claude Opus 5). The authors specified the algorithm and the validation
protocol and directed the implementation through iterative prompting. The resulting code was compared  against that from exact diagonalization. The authors take full responsibility for the content of this manuscript.


\section{Results}
\label{sec:Results}

We now apply SKQD to the impurity Green’s function of the SIAM, represented as a one-dimensional chain of length $L$. Site 0 is the impurity, which carries the Hubbard repulsion $U$; sites $1,\ldots,L-1$ are bath sites with hoppings $t_i$ taken from a converged DMFT bath fit, so the parameters are those of an actual DMFT calculation. We did not solve the DMFT self-consistency equations ourselves; we obtained a converged solution by the method in ref. \cite{Rangi2026} and extracted the resulting bath parameters. We list the parameters in appendix \ref{app:bath}. The impurity level is fixed at $\epsilon_d = -U/2$, so the model is particle–hole symmetric and at half-filling. Because the bath parameters depend on $U$, the hopping parameters and $U$ are chosen together and are not independent knobs \cite{Rangi2026}.

The main quantity throughout is the impurity spectral function
\begin{equation}A(\omega) \;=\; -\frac{1}{\pi}\,\mathrm{Im}\!\left[G^{+}(\omega) + G^{-}(\omega)\right],\end{equation}
with the addition and removal branches evaluated by a Lanczos continued fraction
\begin{equation}G^{+}(\omega) = \langle\chi_{+}|\big[(\omega + E_0 + i\eta) - H\big]^{-1}|\chi_{+}\rangle,\end{equation}
\begin{equation} G^{-}(\omega) = -\langle\chi_{-}|\big[(E_0 - \omega - i\eta) - H\big]^{-1}|\chi_{-}\rangle,\end{equation}
where $|\chi_{+}\rangle = c^{\dagger}_{0\uparrow}|\Psi_0\rangle$ lives in the
$(N{+}1)$ sector given by $\mathcal A^+$, $|\chi_{-}\rangle = c_{0\uparrow}|\Psi_0\rangle$ in the
$(N{-}1)$ sector given by $\mathcal A^-$, and $\eta$ is the Lorentzian broadening or damping factor which is set to $0.1$ for all calculations. All curves labelled "exact (ED)" are in the untruncated sector; every other curve is a SKQD approximation. We do not enforce the particle-hole symmetry in the calculation, the symmetry of the spectra  about $\omega = 0$ serves as a check for the calculation.

\begin{figure*}
    \centering
    \begin{tikzpicture}
      \node[anchor=south west,inner sep=0] (img) at (0,0)
        {\includegraphics[width=0.75\linewidth]{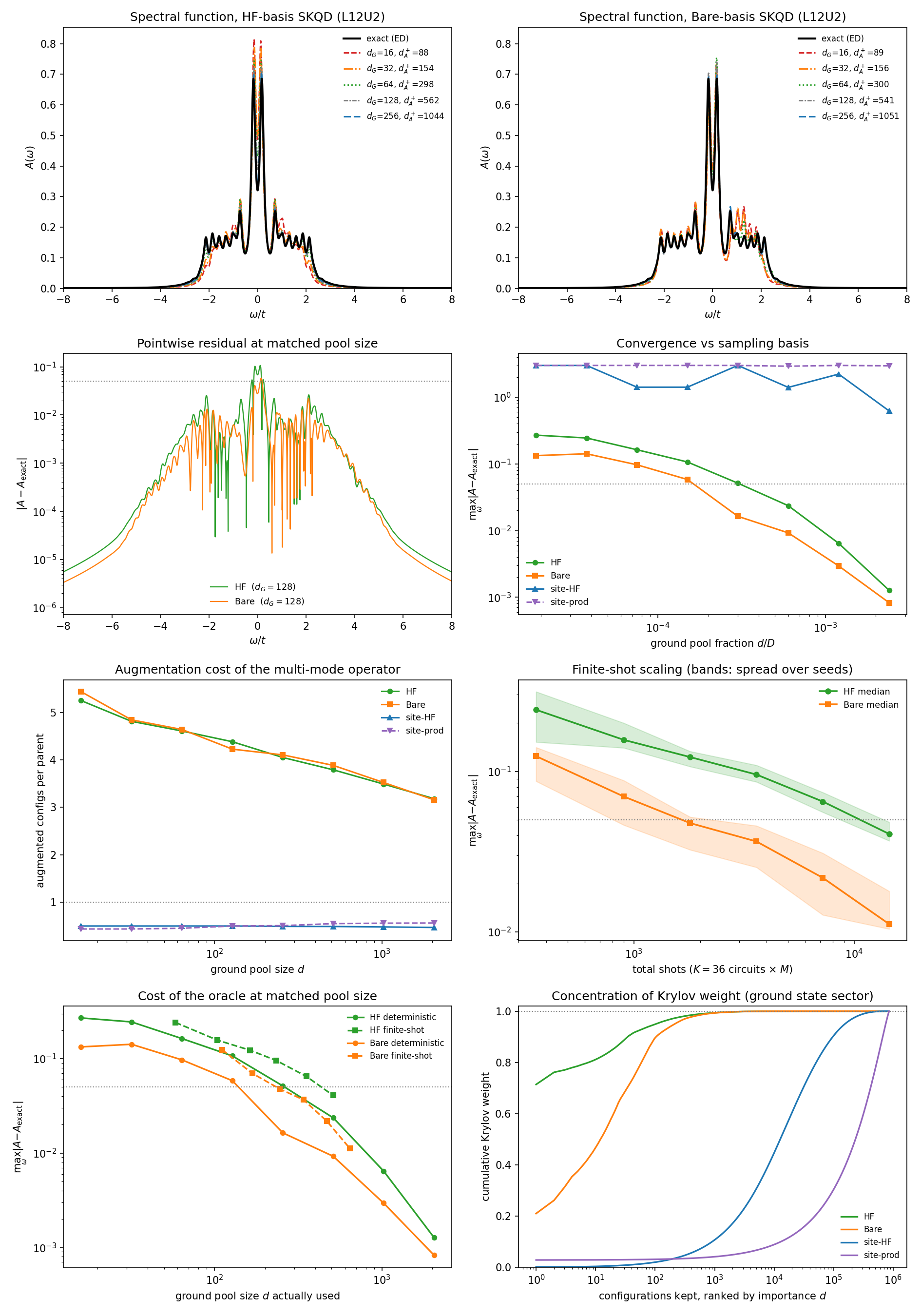}};
      \begin{scope}[x={(img.south east)},y={(img.north west)},
                    every node/.style={anchor=west,font=\small}]
        \node at (0.02,0.985) {(a)};  \node at (0.52,0.985) {(b)};
        \node at (0.02,0.740) {(c)};  \node at (0.52,0.740) {(d)};
        \node at (0.02,0.495) {(e)};  \node at (0.52,0.495) {(f)};
        \node at (0.02,0.250) {(g)};  \node at (0.52,0.250) {(h)};
      \end{scope}
    \end{tikzpicture}
   \caption{SKQD results for the particle--hole-symmetric SIAM chain at
$L=12$, $U=2$. (a),(b) Impurity spectral function $A(\omega)$
reconstructed in the HF and bare orbital bases at increasing subspace size,
against the exact (ED) result. (c) Pointwise error $|A-A_{\rm exact}|$ for
the two bases at matched ground pool size $128$. (d) Error
$\max_\omega|A-A_{\rm exact}|$ versus the fraction $d/D$ of the ground
sector retained; the dotted line is the convergence threshold, $0.05$.
(e) Augmentation cost: configurations added to the branch pool per
ground-pool configuration on applying the multi-mode impurity operator
$c_0=\sum_p w_p d_p$; the dotted line marks one child per parent.
(f) Error versus total shot count, with bands spanning the spread over
random seeds. (g) Error versus the pool size actually realized, comparing deterministic
(exact-amplitude) ranking, the oracle, with finite-shot sampling in each orbital basis. Threshold crossings give the pool size
needed for a target accuracy. (h) Cumulative Krylov weight of the ground sector versus configurations retained, ranked by
importance.}
\label{fig:L12U2}

\end{figure*}
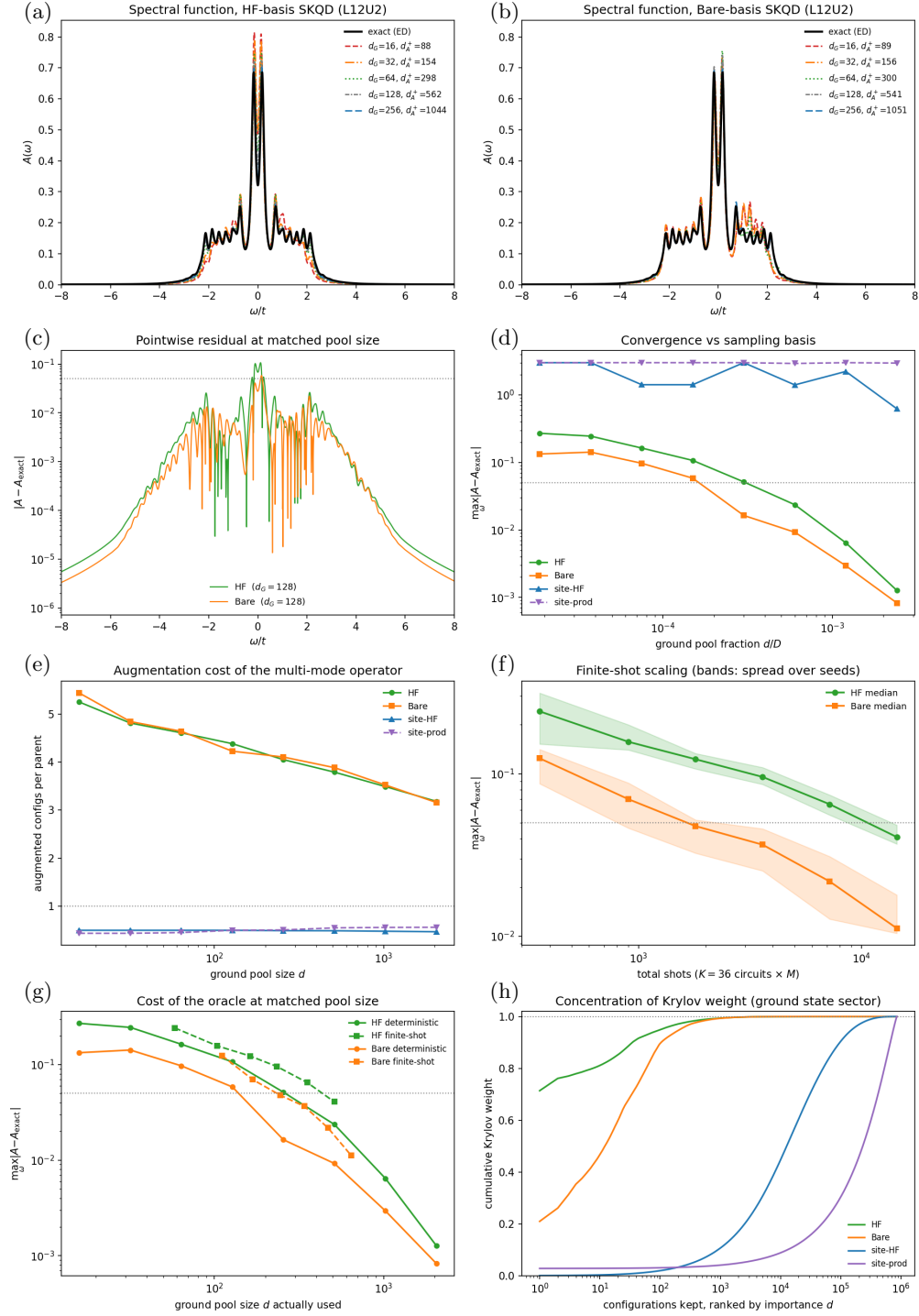

\begin{figure*}
    \centering
    \begin{tikzpicture}
      \node[anchor=south west,inner sep=0] (img) at (0,0)
        {\includegraphics[width=0.75\linewidth]{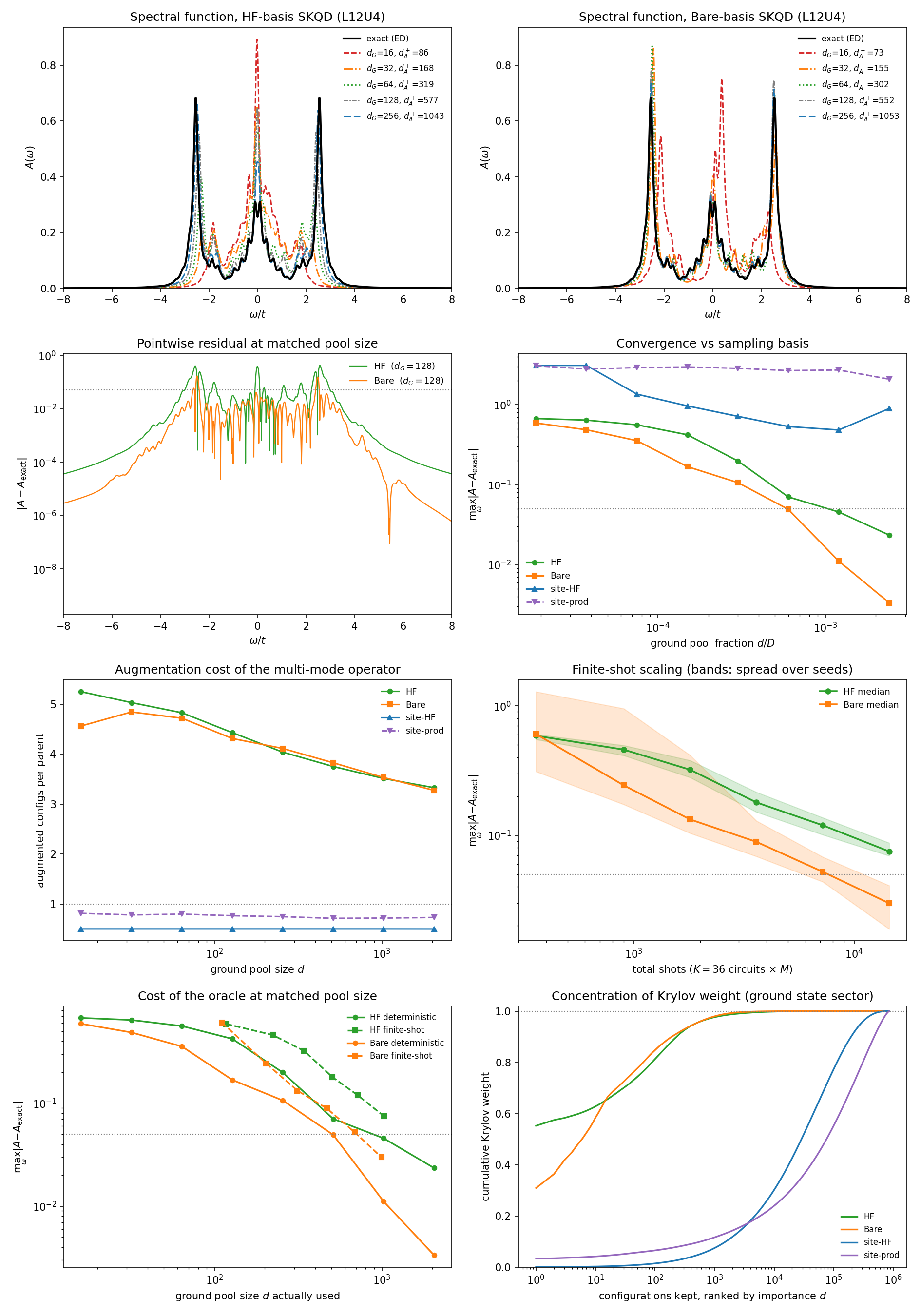}};
      \begin{scope}[x={(img.south east)},y={(img.north west)},
                    every node/.style={anchor=west,font=\small}]
        \node at (0.02,0.985) {(a)};  \node at (0.52,0.985) {(b)};
        \node at (0.02,0.740) {(c)};  \node at (0.52,0.740) {(d)};
        \node at (0.02,0.495) {(e)};  \node at (0.52,0.495) {(f)};
        \node at (0.02,0.250) {(g)};  \node at (0.52,0.250) {(h)};
      \end{scope}
    \end{tikzpicture}
   \caption{SKQD results for the particle--hole-symmetric SIAM chain at
$L=12$, $U=4$. (a),(b) Impurity spectral function $A(\omega)$
reconstructed in the HF and bare orbital bases at increasing subspace size,
against the exact (ED) result. (c) Pointwise error $|A-A_{\rm exact}|$ for
the two bases at matched ground pool size $128$. (d) Error
$\max_\omega|A-A_{\rm exact}|$ versus the fraction $d/D$ of the ground
sector retained; the dotted line is the convergence threshold, $0.05$.
(e) Augmentation cost: configurations added to the branch pool per
ground-pool configuration on applying the multi-mode impurity operator
$c_0=\sum_p w_p d_p$; the dotted line marks one child per parent.
(f) Error versus total shot count, with bands spanning the spread over
random seeds. (g) Error versus the pool size actually realized, comparing deterministic
(exact-amplitude) ranking, the oracle, with finite-shot sampling in each orbital basis. Threshold crossings give the pool size
needed for a target accuracy. (h) Cumulative Krylov weight of the ground sector versus configurations retained, ranked by
importance.}
\label{fig:L12U4}

\end{figure*}

\begin{figure*}
    \centering
    \begin{tikzpicture}
      \node[anchor=south west,inner sep=0] (img) at (0,0)
        {\includegraphics[width=0.75\linewidth]{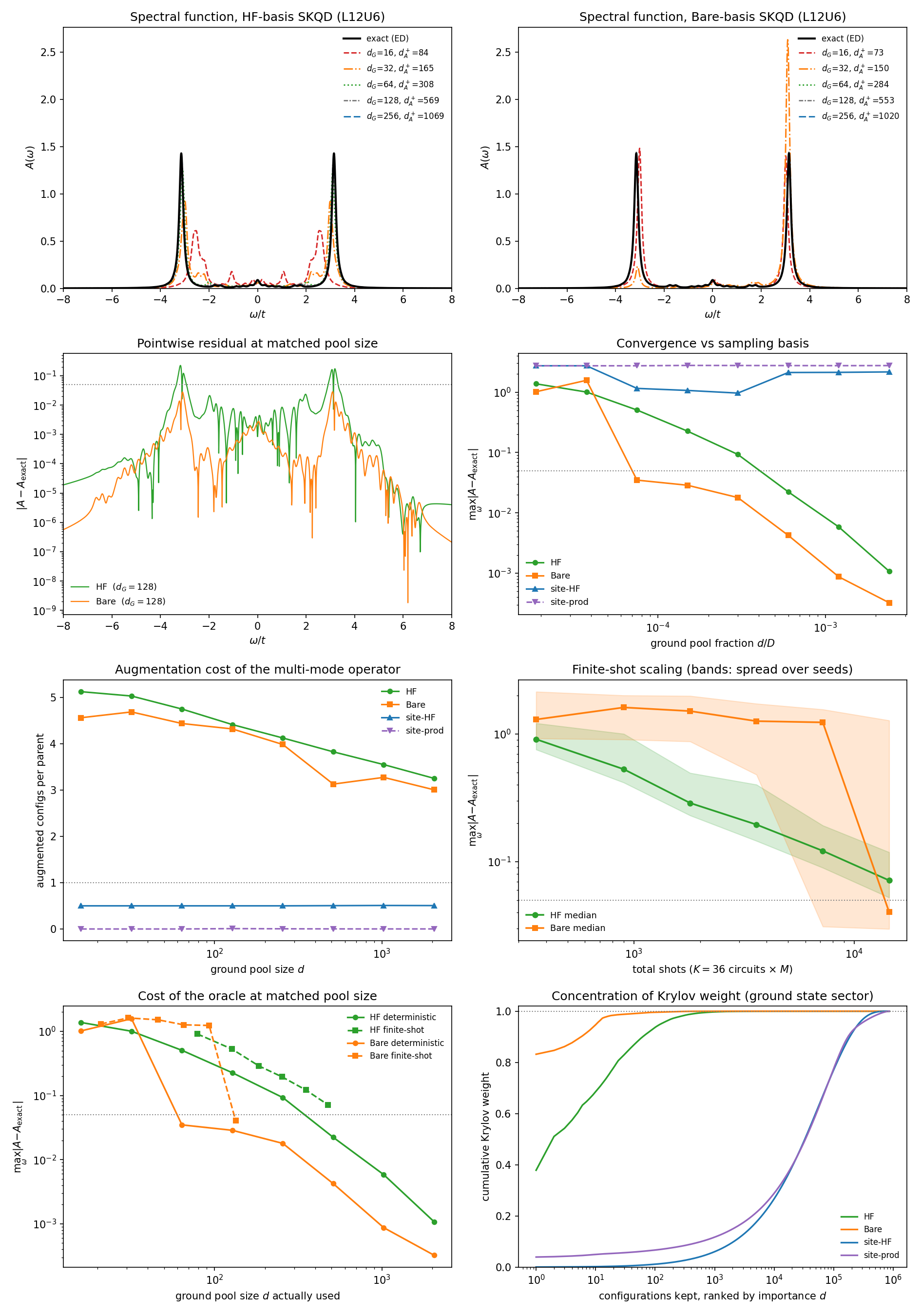}};
      \begin{scope}[x={(img.south east)},y={(img.north west)},
                    every node/.style={anchor=west,font=\small}]
        \node at (0.02,0.985) {(a)};  \node at (0.52,0.985) {(b)};
        \node at (0.02,0.740) {(c)};  \node at (0.52,0.740) {(d)};
        \node at (0.02,0.495) {(e)};  \node at (0.52,0.495) {(f)};
        \node at (0.02,0.250) {(g)};  \node at (0.52,0.250) {(h)};
      \end{scope}
    \end{tikzpicture}
   \caption{SKQD results for the particle--hole-symmetric SIAM chain at
$L=12$, $U=6$. (a),(b) Impurity spectral function $A(\omega)$
reconstructed in the HF and bare orbital bases at increasing subspace size,
against the exact (ED) result. (c) Pointwise error $|A-A_{\rm exact}|$ for
the two bases at matched ground pool size $128$. (d) Error
$\max_\omega|A-A_{\rm exact}|$ versus the fraction $d/D$ of the ground
sector retained; the dotted line is the convergence threshold, $0.05$.
(e) Augmentation cost: configurations added to the branch pool per
ground-pool configuration on applying the multi-mode impurity operator
$c_0=\sum_p w_p d_p$; the dotted line marks one child per parent.
(f) Error versus total shot count, with bands spanning the spread over
random seeds. (g) Error versus the pool size actually realized, comparing deterministic
(exact-amplitude) ranking, the oracle, with finite-shot sampling in each orbital basis. Threshold crossings give the pool size
needed for a target accuracy. (h) Cumulative Krylov weight of the ground sector versus configurations retained, ranked by
importance.}
\label{fig:L12U6}

\end{figure*}

\begin{figure*}
    \centering
    \includegraphics[width=0.75\linewidth]{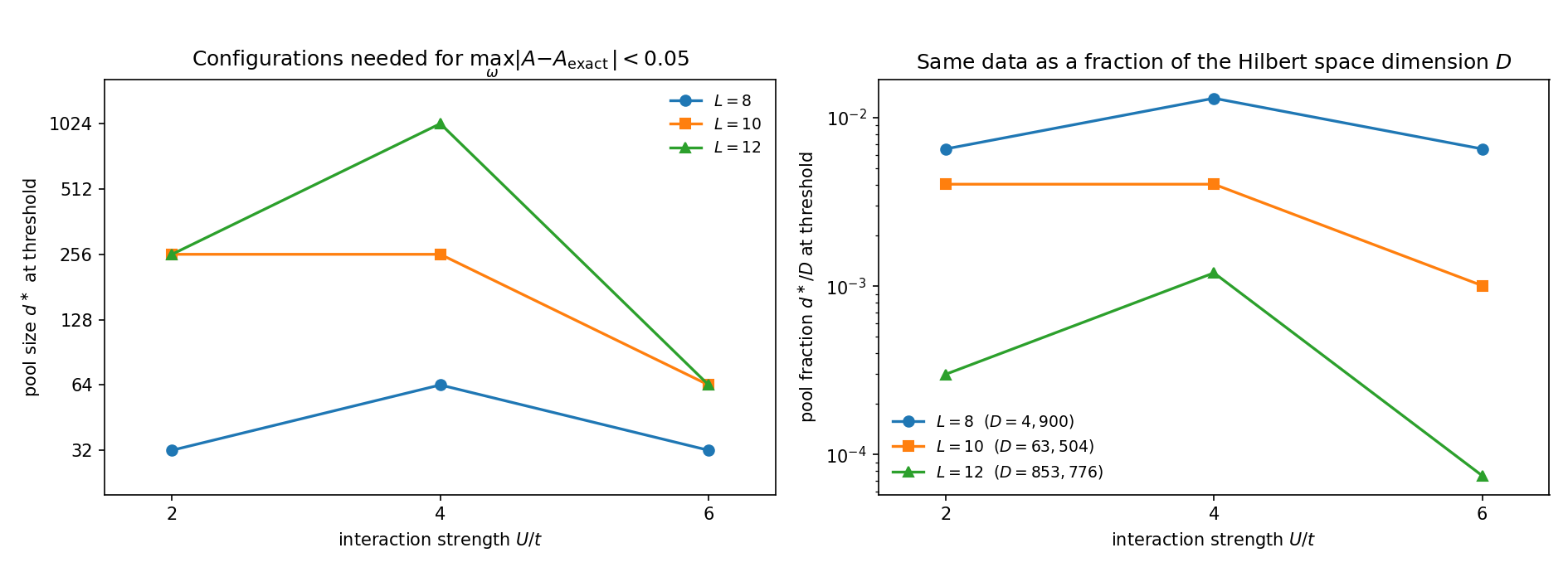}
    \caption{SKQD sampled pool size as a function of interaction strength and system size. Left panel plots the pool size as a function of $U$ for three different system sizes, $L=8, 10,$ and $12$. Right panel plots the ratios between the pool size and the Hilbert space dimension.}
    \label{fig:comparison}
\end{figure*}

SKQD builds the subspace as follows: 
\begin{itemize}
\item Prepare a mean-field or non-interacting bare reference state $|\phi_{ref}\rangle$.

\item Evolve it in real time using the first-order Trotter approximation, $|\phi_k\rangle = \prod_{j=0,1,\cdots k} e^{-iH_U\Delta t} e^{-iH_0\Delta t}|\phi_{ref}\rangle$ for $k = 0,\ldots,K-1$, where $H_U$ is the Hubbard interaction term for the impurity site and $H_0$ is the rest of the Hamiltonian.

\item Measure each $|\phi_k\rangle$ in the computational basis. The union of the sampled bitstrings defines a small subspace. We fix $K=36$ and $\Delta t = \pi / (2U)$.
\item Project $H$ into that subspace of $\mathcal{S}$ and diagonalize classically to obtain the ground state. 
\item Construct the subspaces for $N+1$ and $N-1$ particle sectors $\mathcal A^+$ and $\mathcal A^-$ respectively. 
\item Form the $|\chi^+\rangle$ and $|\chi^-\rangle$ for the initial state to be used for the Lanczos iteration.
\item Use the Lanczos method to find the Green's function.
\end{itemize}

As we use a classical simulator for the calculation, we can obtain the weights of the basis by either deterministic enumeration or stochastic sampling. For the exact enumeration, we rank the configurations by the count of $b$ defined as  \begin{equation}\mathrm{count}(b) \;=\; \sum_{k=0}^{K-1} \big|\langle b | \phi_k\rangle\big|^2\end{equation} and keep the $d$ configurations with the highest counts. Note that $b$ here denotes the bitstring of the basis. For stochastic sampling with a finite number of shots, we draw $M$ samples from each $|\phi_k\rangle$ according to the exact Born probabilities and keep every configuration that is hit at least once. The deterministic version is the infinite-shot limit, while the sampling method is what quantum hardware would deliver.

We consider the system with 1 impurity site and $11$ bath sites with interaction strength $U=2,4,6$. The Hilbert space dimension for half-filling is $(12!/(6! 6!))^2=853776$. We mainly focus on two reference initial states: 1. HF basis from eigenvectors of the chain with the Hartree shift applied, $h_{00} = \epsilon_d + U/2 = 0$. The data of this basis is labeled as "HF" in the figures.
2. Non-interacting bare eigenvectors of the bare chain, $h_{00} = \epsilon_d = -U/2$. The data of this basis is labeled as "Bare" in the figures. And two more reference initial states for comparison: 
3. Real-space HF site basis, given by $W=I$ with the Hubbard term decoupled at the mean-field level. This is the same physical mean-field state, but now a dense superposition. The data of this basis is labeled as "site-HF" in the figures.
4. Site-product basis, $W = I$, a single site configuration as a genuine product state. The data of this basis is labeled as "site-prod" in the figures.

We show results of the interaction strengths  $U=2, 4,$ and $6$ in Figs. 1, 2, and 3 respectively. In Figs. 1,2, and 3, Row 1 left and right panels (panel (a) and panel (b)), we plot the spectral function denoted as $A(\omega)$ over $\omega/t \in [-8, 8]$ in the HF basis and the bare basis respectively. The bold black curve is the exact result; the five colored curves are SKQD reconstructions at increasing subspace pool size, the curves are labelled by $d_G$ (the ground state basis pool size for the space $\mathcal{G}$), and $d_{{A}^{+}}$ (the augmented basis pool size for the space $\mathcal A^+$). Both panels use the deterministic ranking to pick the states to the pool according to their Born weights.

In Figs. 1, 2, and 3, Row 2, the left panel (panel (c)) plots the $|A(\omega) - A_{\rm exact}(\omega)|$ on a logarithmic scale, for the two basis sets at the same ground state pool size fixed at a rather small $d_{\mathcal G}=128$.

In Figs. 1, 2, and 3 Row 2, the right panel (panel (d)) plots the error defined as $\max_\omega |A(\omega) - A_{\rm exact}(\omega)|$ against the ground state pool fraction $d/D$, where $d$ and $D$ are the SKQD and the full ground-sector Hilbert space dimensions respectively, on log–log axes, one curve per basis set. The dotted line is the convergence threshold set at $0.05$. This is the central quantitative comparison which addresses the question of how small a subspace each basis needs to reach a given accuracy. The two site-basis curves stay far above threshold across the whole swept range, demonstrating that they fail as basis sets for SKQD.

In Figs. 1, 2, and 3 Row 3, the left panel (panel (e)) plots the augmentation factor of the pool for the $N+1$ sector: the number of configurations added to the addition-branch pool per ground-pool configuration, versus ground-pool size. The dotted line marks one child per parent. In the site basis $c_{0\uparrow}^\dagger$ touches a single mode, so each parent spawns at most one child and the curves sit at or below $1$. Recall that in a rotated basis $c_{0\uparrow}^\dagger = \sum_p w_p d_{p\uparrow}^\dagger$ is multi-mode and a parent can spawn up to $L - N_{\uparrow}$ children, all of which must enter the branch pool for the moment guarantee to hold. This panel is therefore the price paid for the faster convergence of the rotated basis sets. The ratio decreases with pool size because children of different parents increasingly coincide. For the real-space basis sets, the basis dimension could decrease as some basis states can be destroyed by $c_{0\uparrow}^{\dagger}$. We also note that the effect of including $\mathcal S_{\mathcal A}$ and $\mathcal S_{\mathcal R}$ in $\mathcal A^+$ and $\mathcal A^-$ respectively can be very small for the rotated bases. We find that nearly all of the sampled configurations in $\mathcal S_{\mathcal A}$ and $\mathcal S_{\mathcal R}$ are already contained in $c_{0\uparrow}^{\dagger}\mathcal S_{\mathcal G}$ and $c_{0\uparrow}\mathcal S_{\mathcal G}$. This is a consequence of the multi-mode structure of $c_{0,\uparrow}$ in a rotated basis, which is also the origin of the augmentation cost shown here. It does not hold in the site bases, where $c_{0\uparrow}$ touches a single mode only.

In Figs. 1, 2, and 3 Row 3, the right panel (panel (f)) plots the error against the total shot count $K \times M$, for the two different basis sets "HF" and "Bare". Solid lines are the median over independent random seeds; the shaded bands span the minimum and maximum observed among 20 runs across those seeds. All sampling uses exact Born probabilities, so this ignores noise from any hardware error. The bands are just range statistics over a small number of seeds, not confidence intervals, given that we only have 20 runs. A wide band indicates that the outcome depends strongly on whether a particular seed happened to sample the few configurations that carry the spectral weight.

In Figs. 1, 2, and 3 Row 4, left panel (panel (g)) plots the error against the number of configurations actually obtained, comparing the
deterministic ranking with finite-shot sampling for each orbital basis in its own color. $K \times M$ shots yield far fewer than $K \times M$ distinct bitstrings because of collisions, and the collision rate depends on how sharply peaked the Born distributions are, so a basis with highly concentrated weight converts shots into unique configurations much less efficiently than a diffuse one.

In Figs. 1, 2, and 3 Row 4, right panel (panel (h)) plots the cumulative normalized importance, $\sum_{n \le d} \mathrm{count}(n) \big/ \sum_{n} \mathrm{count}(n)$, with configurations ordered by decreasing importance, against $d$ on a logarithmic axis, for all four basis sets. It measures how sharply the Krylov space concentrates on the full Hilbert space, independent of any Green's function evaluation. A curve that rises steeply and saturates early identifies a basis in which few configurations suffice; one that rises slowly identifies a "delocalized" problem that no small subspace can capture.

The applicability of the proposed method clearly depends on how the required resources scale with the number of bath sites. Additional results for $L=10$ and $L=8$ are given in Appendix \ref{app:L8L10}. Among the possible comparisons, a crucial indicator of viability on quantum hardware is whether the sampled pool size grows exponentially or more modestly than the Hilbert-space dimension.

In Fig. 4 we plot the comparison of the sampled pool size required to get the error in the spectral function as defined above to be less than $0.05$. Left panel plots the pool size as a function of $U$ for three different system sizes, $L=8, 10,$ and $12$. There is an increase in the pool size as the interaction strength goes from weak coupling ($U=2$) to intermediate coupling ($U=4$), and then a decrease at strong coupling ($U=6$) in which the gap becomes more pronounced in the spectral function. The pool size required to reach the threshold is non-monotonic in $U$, peaking at intermediate coupling for all three system sizes. At weak coupling the ground state is close to the non-interacting Fermi sea, which is a single determinant; at strong coupling it approaches a local moment on the impurity times a filled bath, which is again nearly a single determinant once the basis contains a near-atomic impurity orbital. At intermediate coupling the state is neither, and the spectral function carries coherent quasiparticle weight and incoherent Hubbard weight simultaneously, so a larger set of configurations is needed to resolve both. The requirement is therefore largest precisely in the crossover region.

It is also instructive to compare the pool size with the dimension of the Hilbert space. The right panel in Fig. \ref{fig:comparison} shows the pool size required at threshold as a fraction of the Hilbert space dimension. This fraction decreases monotonically with system size at every interaction strength considered, and the decrease is substantial across the accessible range. The subspace needed to reach a fixed spectral accuracy occupies a progressively smaller portion of the Hilbert space as the problem grows. Three system sizes do not permit a controlled finite-size scaling analysis, and we do not attempt one here. The trend is nevertheless consistent at weak, intermediate, and strong coupling alike, which is the behavior a sampling-based subspace method must exhibit if it is to remain useful at larger bath discretizations.

\section{Conclusion}
\label{sec:Conclusion}
We have introduced a formalism for computing the single-particle Green's function within the sample-based Krylov quantum diagonalization framework. By sampling the Krylov subspaces generated by short-time unitary evolution and evaluating all matrix elements classically, the method reconstructs the Lehmann representation (or an equivalent Lanczos continued fraction) entirely inside a truncated Hilbert space. Applied to the particle-hole-symmetric single-impurity Anderson model (SIAM) with a discrete bath taken from a converged dynamical mean-field theory (DMFT) calculation, the approach reproduces the expected interaction-driven evolution of the impurity spectral function across the metal-insulator transition. Importantly, good accuracy is already obtained with a subspace that comprises only a small fraction of the full Hilbert-space dimension, making the algorithm a realistic candidate for near-term quantum hardware. For $L=12$ it reduces the Hilbert space by about three to four orders of magnitude while still giving a reasonable approximation to the spectral function. How the required subspace size scales with the number of bath sites remains the central open question. The trend as shown in Fig. \ref{fig:comparison} is encouraging, but three system sizes do not permit a controlled extrapolation, and larger chains are beyond the reach of the classical simulation used here without a substantial amount of work on optimizing the simulation.

Given the importance of DMFT in the study of strongly correlated materials \cite{Kotliar2006}, a considerable amount of effort has already been devoted to the design and experimental demonstration of quantum algorithms for DMFT impurity solvers. Hybrid quantum-classical schemes for evaluating impurity Green's functions have been realized on superconducting and trapped-ion platforms, and DFT+DMFT workflows for real materials have begun to appear \cite{Baul2023,Singh2026,Rangi2026,Keen2020,Steckmann2023,Selisko2025,Kemper2025,Endo2020,Rizzo2022,Jamet2021,Jamet2022,GreeneDiniz2024,Keen2021,Bishop2023}. A comprehensive recent review of the field is given by Ayral \cite{Ayral2025}. The SKQD construction presented here offers a complementary, ancilla-free, sampling-based route to the real-frequency spectral function and could therefore serve as an alternative impurity solver for the most expensive step of the DMFT self-consistency loop.

Most ingredients of the algorithm are generic and can be transferred to a broad class of quantum many-body problems in which excitation spectra or dynamical correlators are required. The principal practical challenge arises when the quadratic part of the Hamiltonian is diagonalized (the ``bare'' or Hartree-Fock orbital basis). In that representation a local interaction is transformed into a "non-local" interaction. For the SIAM, only a single site carries the Hubbard term, so the extra cost remains modest. The local density operator is a rank-one outer product and the necessary time evolution can still be realized with a linear number of Givens rotations. For lattice models with interactions on every site (e.g., the Hubbard model) the same transformation produces long-range terms whose circuit depth grows more rapidly. Realistic multi-orbital calculations within DMFT, which typically involve Kanamori interactions and non-diagonal spin-orbit couplings, will require correspondingly more elaborate fermionic decompositions.

Another extension of the present framework is the evaluation of two-particle Green’s functions and the associated local vertex functions. Methods that go beyond single-site DMFT through diagrammatic expansions such as the parquet approximation, dynamical vertex approximation, and dual-fermion approach \cite{Yang2011,Toschi2007,Rubtsov2008,Rohringer2018,Fotso2022} all rely on a perturbative expansion around a DMFT solution and therefore require accurate local two-particle vertices. By constructing and sampling Krylov subspaces generated from two-particle seed states, the sample-based approach developed here could in principle supply these vertices. Whether the favorable basis compression observed for the one-particle spectral function persists at the two-particle level remains an open question that we leave for future investigation.

We note that a closely related sample-based approach to dynamical spectral functions, constructed directly from bitstring-sampled subspaces, has appeared very recently \cite{BonillaVargas2026}. That work also discusses connections between SKQD-type methods and AI and demonstrates applications to molecular systems.

\section{Acknowledgements} 
We thank Mohommed Rahman for useful discussions. This manuscript is based on work supported by the National Science Foundation under awards OAC-2150491 and OAC-2447810 with additional support from the Center for Computation and Technology at Louisiana State University. This work used high-performance computational resources provided by the Louisiana Optical Network Initiative and HPC@LSU computing.

\appendix

\section{Lanczos method}
\label{app:lanczos}

We collect here the details of the Lanczos construction used to evaluate the
continued fraction of Eq. \ref{eq:Lanczos_CF} inside the truncated space,
largely following Lin and Gubernatis~\cite{Lin1993}. The recursion is carried
out entirely within the augmented subspaces $\mathcal{A}^{\pm}$ of
Eq.~(\ref{eq:Aspace}).

Let $V$ denote the isometry whose columns are the configurations in
$\mathcal{A}^{\pm}$, so that $V^{\dagger}V=1$ and
$\hat{P}_{\mathcal{A}^{\pm}}=VV^{\dagger}$ is the orthogonal projector onto
$\mathcal{A}^{\pm}$. The recursion is applied to the compressed Hamiltonian
\begin{equation}
H_{\mathcal{A}^{\pm}}\equiv V^{\dagger}HV ,
\label{eq:compressed}
\end{equation}
a matrix of dimension $d_{\mathcal{A}^{\pm}}=|\mathcal{A}^{\pm}|$ rather than
the full $N\pm1$ sector dimension; it is the principal submatrix of $H$
obtained by retaining only the rows and columns labelled by configurations in
$\mathcal{A}^{\pm}$. Viewed in the full sector the same operator reads
$\hat{P}_{\mathcal{A}^{\pm}}H\hat{P}_{\mathcal{A}^{\pm}}
=VH_{\mathcal{A}^{\pm}}V^{\dagger}$.

Because $\mathcal{A}^{\pm}$ was constructed to contain
$c^{(\dagger)}_{0\uparrow}\mathcal{S}_{\mathcal{G}}$, the seed satisfies
$\hat{P}_{\mathcal{A}^{\pm}}|\chi^{\pm}\rangle=|\chi^{\pm}\rangle$ exactly, so
no weight is lost in passing from the full sector to the subspace. The
truncation enters only through the action of $H$ on states inside
$\mathcal{A}^{\pm}$: matrix elements connecting $\mathcal{A}^{\pm}$ to its
complement are discarded.

Starting from the normalized seed
\begin{equation}
|u_{0}\rangle=\frac{V^{\dagger}|\chi^{\pm}\rangle}{\|\chi^{\pm}\|},
\qquad |u_{-1}\rangle\equiv 0,\qquad \beta_{0}\equiv 0 ,
\label{eq:seed}
\end{equation}
the three-term recurrence proceeds for $j=0,1,\dots,n_{L}-1$ can be written as \cite{Lin1993}
\begin{align}
\alpha_{j}&=\langle u_{j}|H_{\mathcal{A}^{\pm}}|u_{j}\rangle ,
\label{eq:alpha}\\[2pt]
|r\rangle&=H_{\mathcal{A}^{\pm}}|u_{j}\rangle-\alpha_{j}|u_{j}\rangle
              -\beta_{j}|u_{j-1}\rangle ,
\label{eq:resid}\\[2pt]
\beta_{j+1}&=| |r\rangle| ,
\qquad |u_{j+1}\rangle=|r\rangle/\beta_{j+1} .
\label{eq:beta}
\end{align}
%
The $\{\alpha_{j},\beta_{j}\}$ so generated are the diagonal and off-diagonal entries of the $n_{L}\times n_{L}$ tridiagonal matrix that defines the continued fraction in Eq. \ref{eq:Lanczos_CF}. 

The recursion terminates at $j=n_{L}-1$, or earlier if $\beta_{j+1}$ falls
below a fixed threshold. In exact arithmetic the latter signals that an invariant subspace of $H_{\mathcal{A}^{\pm}}$ containing $|\chi^{\pm}\rangle$
has been exhausted, and the continued fraction is numerically exact within it.
Only three vectors of length $d_{\mathcal{A}^{\pm}}$ are held at any time, so
the storage is $O(d_{\mathcal{A}^{\pm}})$ and the cost is $n_{L}$ sparse
matrix-vector products.

Since the poles are broadened by $\eta$, the recursion needs to resolve the
spectrum only on that scale. Further Lanczos steps split already-broadened
structure into features narrower than $\eta$, which the Lorentzian convolution
then washes out. It is therefore not necessary to take
$n_{L}$ comparable to $d_{\mathcal{A}^{\pm}}$; we cap $n_{L}$ below $250$ for
all SKQD reconstructions.

In practice $H_{\mathcal{A}^{\pm}}$ is obtained by restriction of the sparse
Hamiltonian. When even the submatrix is too large to store, the same recursion
is carried out by embedding each Lanczos vector in the full sector, applying
$H$, and restricting the result, that is, by evaluating
$V^{\dagger}HV|u_{j}\rangle$ one vector at a time. The projector
$\hat{P}_{\mathcal{A}^{\pm}}$ is then never formed explicitly.

\section{Discrete bath parameters}
\label{app:bath}

The impurity solvers in the main text use a particle-hole-symmetric
single-impurity Anderson model in chain geometry. The impurity level is
fixed at \(\epsilon_d=-U/2\) and all bath on-site energies vanish,
\(\epsilon_i=0\) for \(i=1,\ldots,L-1\). The nearest-neighbor hoppings
\(t_i\) (\(i=0,\ldots,L-2\)) are taken from a conventional
exact-diagonalization DMFT bath fit at the indicated
interaction~\cite{Rangi2026}. Site \(0\) is the impurity, so \(t_0\) is
the hybridization between the impurity and the first bath site.
Tables~\ref{tab:bath-L8}--\ref{tab:bath-L12} list the numerical values.

\begin{table}[htbp]
\centering
\caption{Hopping parameters \(t_i\) for \(L=8\) (one impurity and seven bath sites).}
\label{tab:bath-L8}
\begin{tabular}{lccc}
\hline\hline
 & \(U=2\) & \(U=4\) & \(U=6\) \\
\hline
\(\epsilon_d\) & \(-1.0\) & \(-2.0\) & \(-3.0\) \\
\hline
\(t_0\) & \(0.6034\) & \(0.4789\) & \(0.2835\) \\
\(t_1\) & \(1.0536\) & \(1.3133\) & \(0.8222\) \\
\(t_2\) & \(0.6668\) & \(0.4745\) & \(0.5096\) \\
\(t_3\) & \(0.7220\) & \(1.0956\) & \(0.6464\) \\
\(t_4\) & \(0.9595\) & \(0.8162\) & \(0.6095\) \\
\(t_5\) & \(0.3628\) & \(0.4744\) & \(0.4026\) \\
\(t_6\) & \(1.0747\) & \(1.2186\) & \(0.6861\) \\
\hline\hline
\end{tabular}
\end{table}

\begin{table}[htbp]
\centering
\caption{Hopping parameters \(t_i\) for \(L=10\) (one impurity and nine bath sites).}
\label{tab:bath-L10}
\begin{tabular}{lccc}
\hline\hline
 & \(U=2\) & \(U=4\) & \(U=6\) \\
\hline
\(\epsilon_d\) & \(-1.0\) & \(-2.0\) & \(-3.0\) \\
\hline
\(t_0\) & \(0.7655\) & \(0.5757\) & \(0.3607\) \\
\(t_1\) & \(0.6695\) & \(0.7560\) & \(0.7264\) \\
\(t_2\) & \(0.8610\) & \(1.3363\) & \(0.9508\) \\
\(t_3\) & \(0.9144\) & \(0.9799\) & \(0.8357\) \\
\(t_4\) & \(0.6713\) & \(0.5900\) & \(0.6241\) \\
\(t_5\) & \(0.5039\) & \(0.4638\) & \(0.4847\) \\
\(t_6\) & \(0.4385\) & \(0.4266\) & \(0.4438\) \\
\(t_7\) & \(0.3787\) & \(0.3680\) & \(0.3777\) \\
\(t_8\) & \(0.2834\) & \(0.2772\) & \(0.2851\) \\
\hline\hline
\end{tabular}
\end{table}

\begin{table}[htbp]
\centering
\caption{Hopping parameters \(t_i\) for \(L=12\) (one impurity and eleven bath sites).}
\label{tab:bath-L12}
\begin{tabular}{lccc}
\hline\hline
 & \(U=2\) & \(U=4\) & \(U=6\) \\
\hline
\(\epsilon_d\) & \(-1.0\) & \(-2.0\) & \(-3.0\) \\
\hline
\(t_0\)    & \(0.7398\) & \(0.6238\) & \(0.4070\) \\
\(t_1\)    & \(1.1067\) & \(0.8938\) & \(1.0182\) \\
\(t_2\)    & \(0.8725\) & \(1.2199\) & \(1.1381\) \\
\(t_3\)    & \(0.9830\) & \(0.8192\) & \(0.8052\) \\
\(t_4\)    & \(0.8786\) & \(0.7500\) & \(0.7806\) \\
\(t_5\)    & \(0.8442\) & \(0.9788\) & \(0.9986\) \\
\(t_6\)    & \(0.9440\) & \(1.0098\) & \(0.9829\) \\
\(t_7\)    & \(0.6609\) & \(0.6571\) & \(0.6327\) \\
\(t_8\)    & \(0.9524\) & \(0.4824\) & \(0.4861\) \\
\(t_9\)    & \(0.4618\) & \(0.4020\) & \(0.4036\) \\
\(t_{10}\) & \(0.8756\) & \(0.3029\) & \(0.3055\) \\
\hline\hline
\end{tabular}
\end{table}

\section{Spectral functions for \(L=10\) and \(L=8\)}
\label{app:L8L10}

\begin{figure*}
    \centering
    \begin{tikzpicture}
      \node[anchor=south west,inner sep=0] (img) at (0,0)
        {\includegraphics[width=0.75\linewidth]{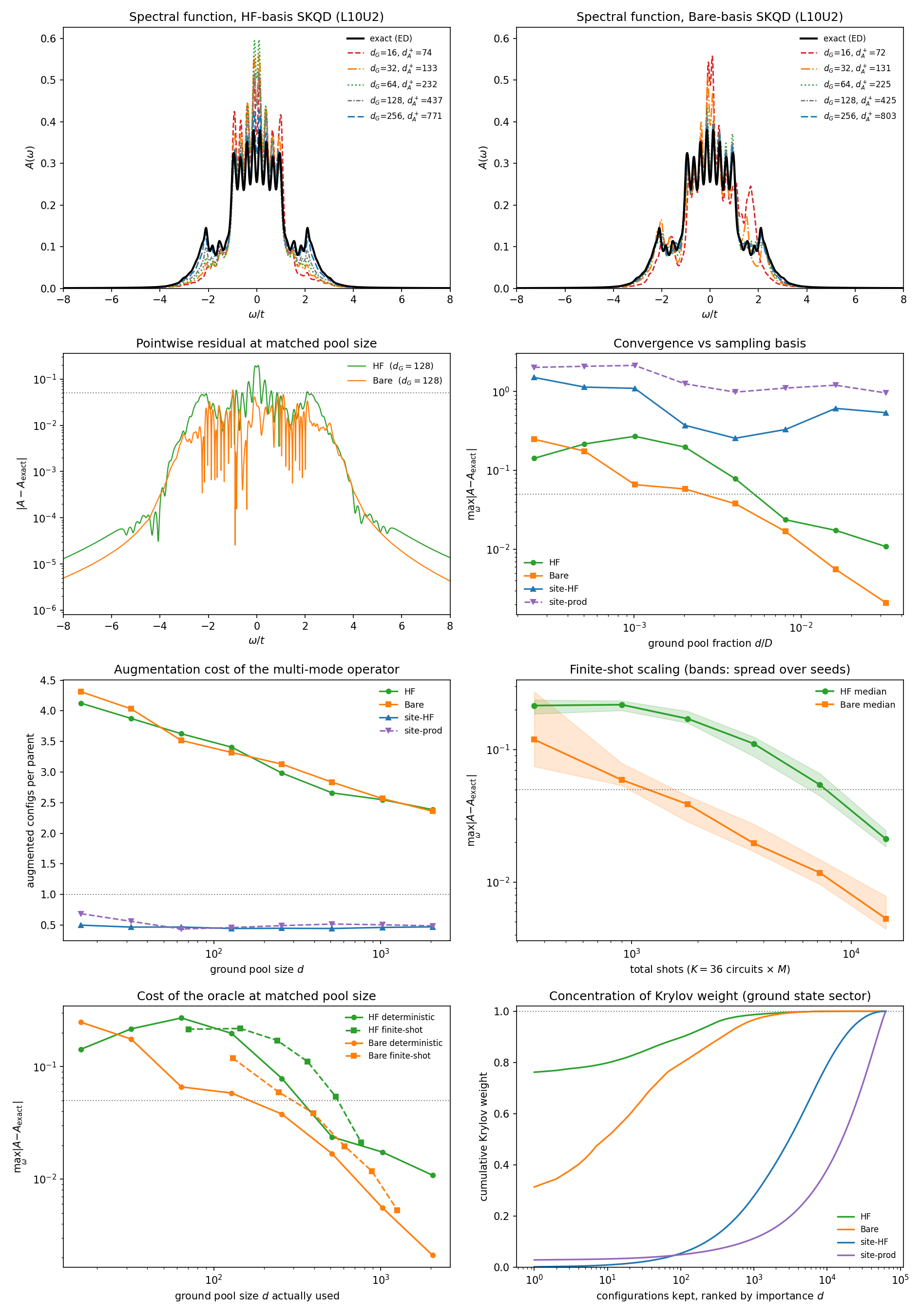}};
      \begin{scope}[x={(img.south east)},y={(img.north west)},
                    every node/.style={anchor=west,font=\small}]
        \node at (0.02,0.985) {(a)};  \node at (0.52,0.985) {(b)};
        \node at (0.02,0.740) {(c)};  \node at (0.52,0.740) {(d)};
        \node at (0.02,0.495) {(e)};  \node at (0.52,0.495) {(f)};
        \node at (0.02,0.250) {(g)};  \node at (0.52,0.250) {(h)};
      \end{scope}
    \end{tikzpicture}
   \caption{SKQD results for the particle--hole-symmetric SIAM chain at
$L=10$, $U=2$. (a),(b) Impurity spectral function $A(\omega)$
reconstructed in the HF and bare orbital bases at increasing subspace size,
against the exact (ED) result. (c) Pointwise error $|A-A_{\rm exact}|$ for
the two bases at matched ground pool size $128$. (d) Error
$\max_\omega|A-A_{\rm exact}|$ versus the fraction $d/D$ of the ground
sector retained; the dotted line is the convergence threshold, $0.05$.
(e) Augmentation cost: configurations added to the branch pool per
ground-pool configuration on applying the multi-mode impurity operator
$c_0=\sum_p w_p d_p$; the dotted line marks one child per parent.
(f) Error versus total shot count, with bands spanning the spread over
random seeds. (g) Error versus the pool size actually realized, comparing deterministic
(exact-amplitude) ranking, the oracle, with finite-shot sampling in each orbital basis. Threshold crossings give the pool size
needed for a target accuracy. (h) Cumulative Krylov weight of the ground sector versus configurations retained, ranked by
importance.}
\label{fig:L10U2}

\end{figure*}
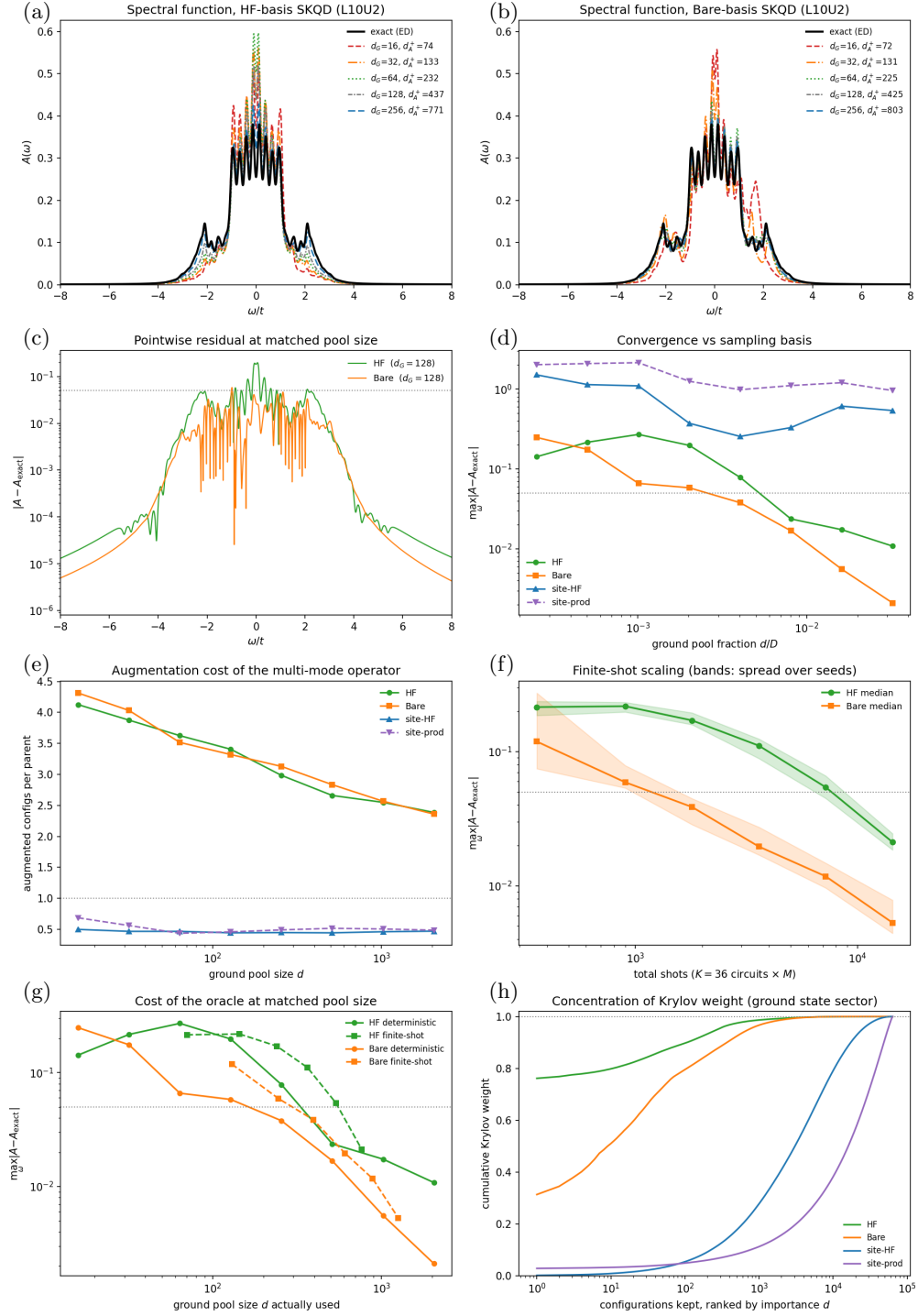

\begin{figure*}
    \centering
    \begin{tikzpicture}
      \node[anchor=south west,inner sep=0] (img) at (0,0)
        {\includegraphics[width=0.75\linewidth]{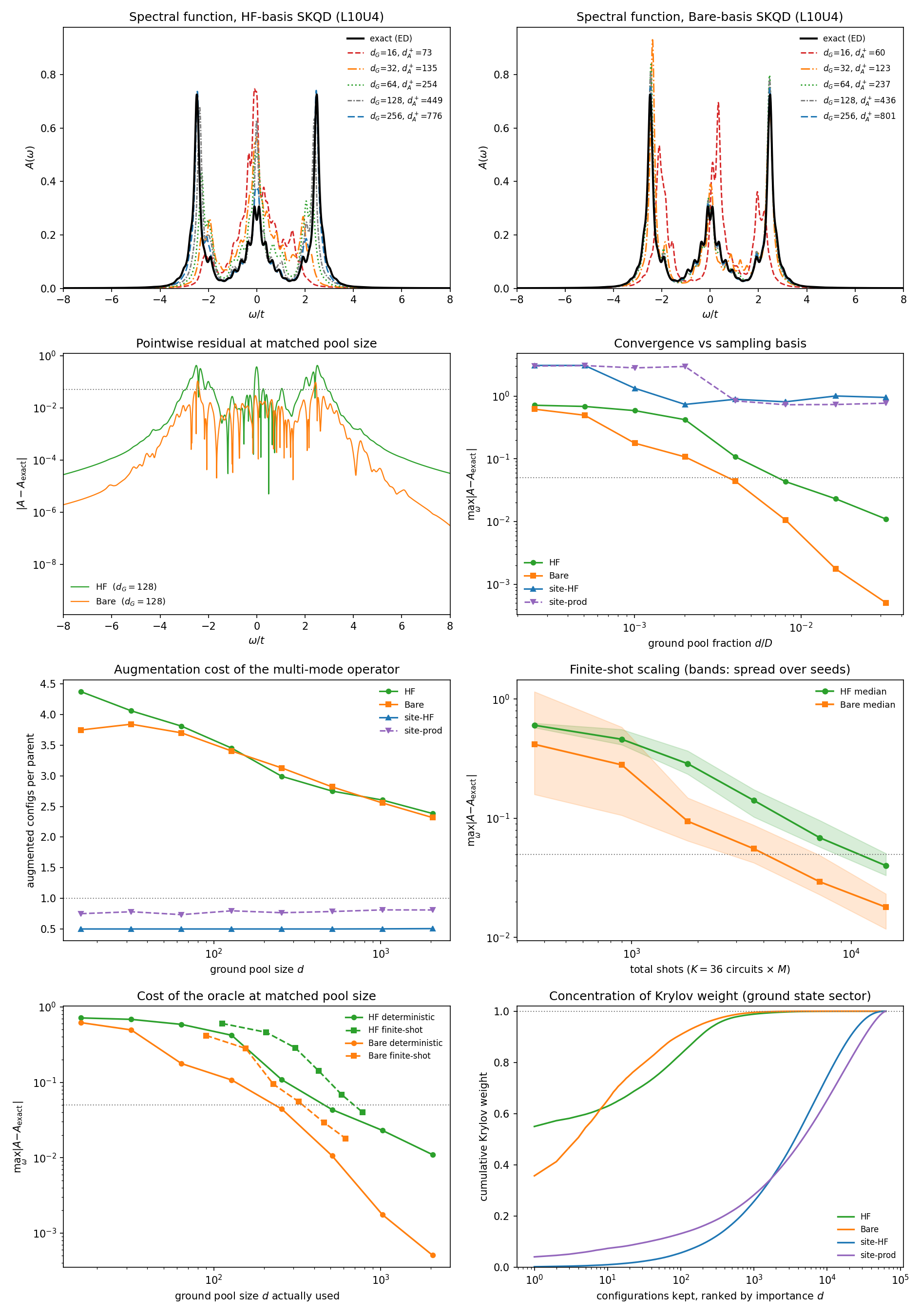}};
      \begin{scope}[x={(img.south east)},y={(img.north west)},
                    every node/.style={anchor=west,font=\small}]
        \node at (0.02,0.985) {(a)};  \node at (0.52,0.985) {(b)};
        \node at (0.02,0.740) {(c)};  \node at (0.52,0.740) {(d)};
        \node at (0.02,0.495) {(e)};  \node at (0.52,0.495) {(f)};
        \node at (0.02,0.250) {(g)};  \node at (0.52,0.250) {(h)};
      \end{scope}
    \end{tikzpicture}
   \caption{SKQD results for the particle--hole-symmetric SIAM chain at
$L=10$, $U=4$. (a),(b) Impurity spectral function $A(\omega)$
reconstructed in the HF and bare orbital bases at increasing subspace size,
against the exact (ED) result. (c) Pointwise error $|A-A_{\rm exact}|$ for
the two bases at matched ground pool size $128$. (d) Error
$\max_\omega|A-A_{\rm exact}|$ versus the fraction $d/D$ of the ground
sector retained; the dotted line is the convergence threshold, $0.05$.
(e) Augmentation cost: configurations added to the branch pool per
ground-pool configuration on applying the multi-mode impurity operator
$c_0=\sum_p w_p d_p$; the dotted line marks one child per parent.
(f) Error versus total shot count, with bands spanning the spread over
random seeds. (g) Error versus the pool size actually realized, comparing deterministic
(exact-amplitude) ranking, the oracle, with finite-shot sampling in each orbital basis. Threshold crossings give the pool size
needed for a target accuracy. (h) Cumulative Krylov weight of the ground sector versus configurations retained, ranked by
importance.}
\label{fig:L10U4}

\end{figure*}

\begin{figure*}
    \centering
    \begin{tikzpicture}
      \node[anchor=south west,inner sep=0] (img) at (0,0)
        {\includegraphics[width=0.75\linewidth]{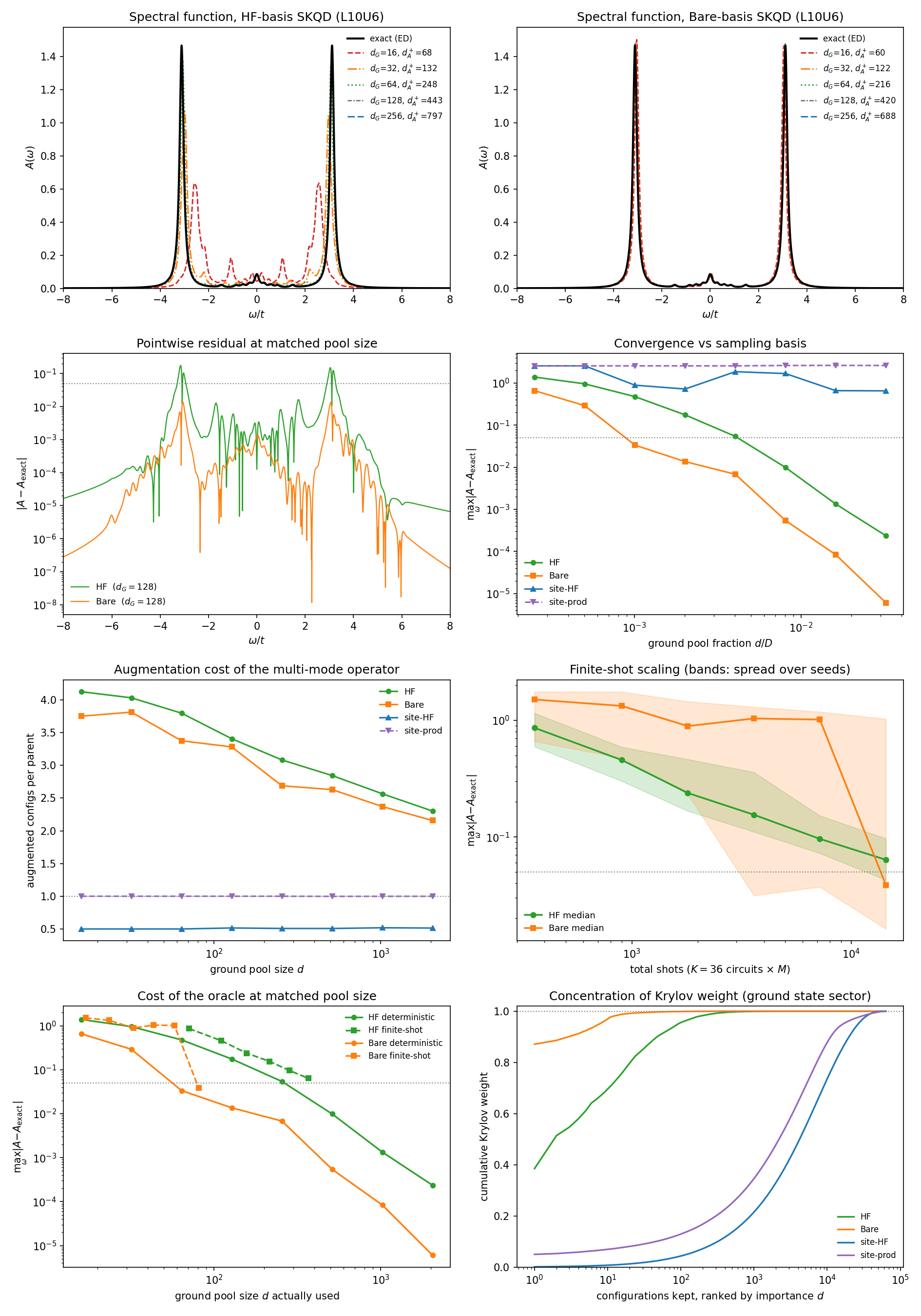}};
      \begin{scope}[x={(img.south east)},y={(img.north west)},
                    every node/.style={anchor=west,font=\small}]
        \node at (0.02,0.985) {(a)};  \node at (0.52,0.985) {(b)};
        \node at (0.02,0.740) {(c)};  \node at (0.52,0.740) {(d)};
        \node at (0.02,0.495) {(e)};  \node at (0.52,0.495) {(f)};
        \node at (0.02,0.250) {(g)};  \node at (0.52,0.250) {(h)};
      \end{scope}
    \end{tikzpicture}
   \caption{SKQD results for the particle--hole-symmetric SIAM chain at
$L=10$, $U=6$. (a),(b) Impurity spectral function $A(\omega)$
reconstructed in the HF and bare orbital bases at increasing subspace size,
against the exact (ED) result. (c) Pointwise error $|A-A_{\rm exact}|$ for
the two bases at matched ground pool size $128$. (d) Error
$\max_\omega|A-A_{\rm exact}|$ versus the fraction $d/D$ of the ground
sector retained; the dotted line is the convergence threshold, $0.05$.
(e) Augmentation cost: configurations added to the branch pool per
ground-pool configuration on applying the multi-mode impurity operator
$c_0=\sum_p w_p d_p$; the dotted line marks one child per parent.
(f) Error versus total shot count, with bands spanning the spread over
random seeds. (g) Error versus the pool size actually realized, comparing deterministic
(exact-amplitude) ranking, the oracle, with finite-shot sampling in each orbital basis. Threshold crossings give the pool size
needed for a target accuracy. (h) Cumulative Krylov weight of the ground sector versus configurations retained, ranked by
importance.}
\label{fig:L10U6}

\end{figure*}

\begin{figure*}
    \centering
    \begin{tikzpicture}
      \node[anchor=south west,inner sep=0] (img) at (0,0)
        {\includegraphics[width=0.75\linewidth]{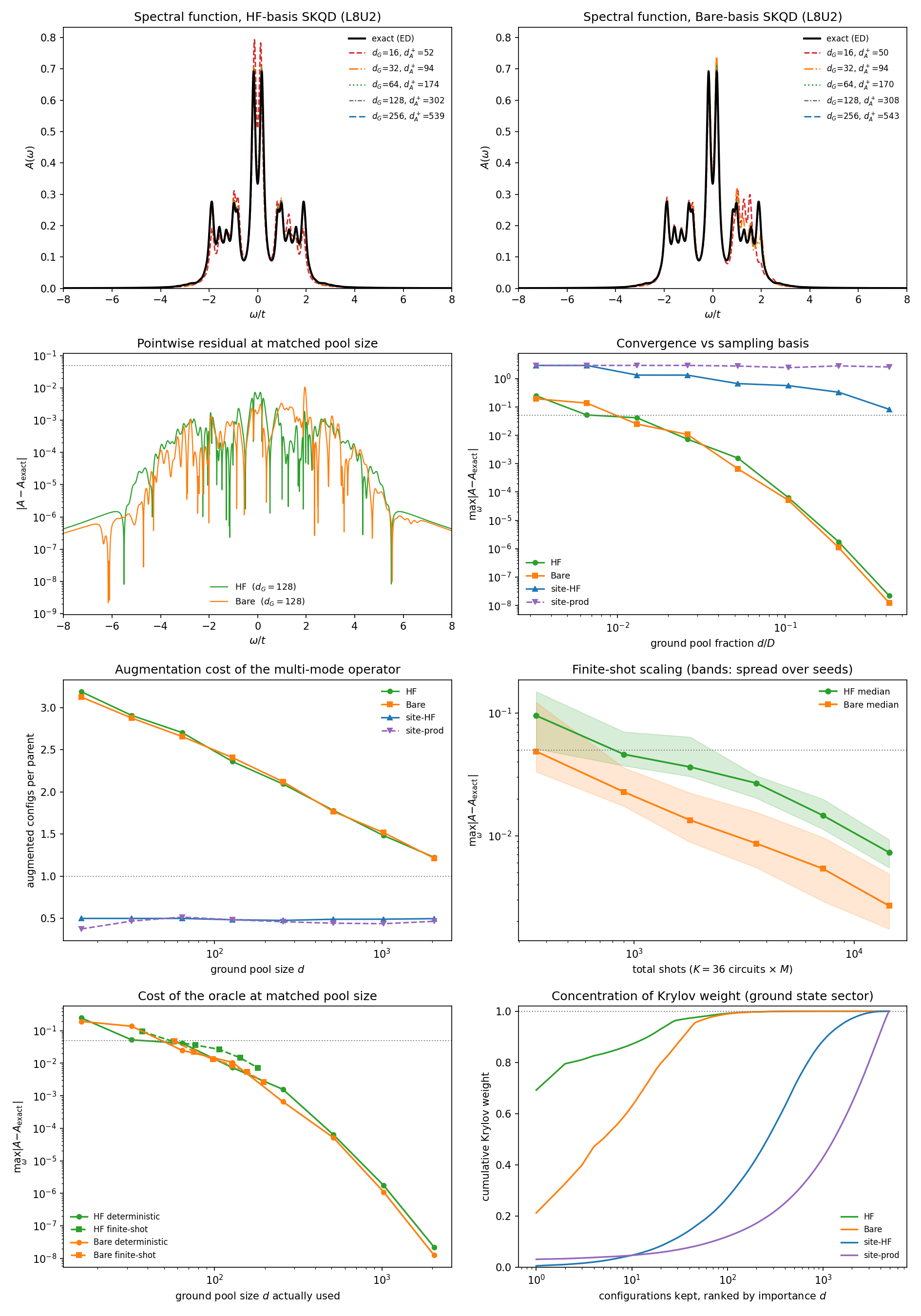}};
      \begin{scope}[x={(img.south east)},y={(img.north west)},
                    every node/.style={anchor=west,font=\small}]
        \node at (0.02,0.985) {(a)};  \node at (0.52,0.985) {(b)};
        \node at (0.02,0.740) {(c)};  \node at (0.52,0.740) {(d)};
        \node at (0.02,0.495) {(e)};  \node at (0.52,0.495) {(f)};
        \node at (0.02,0.250) {(g)};  \node at (0.52,0.250) {(h)};
      \end{scope}
    \end{tikzpicture}
   \caption{SKQD results for the particle--hole-symmetric SIAM chain at
$L=8$, $U=2$. (a),(b) Impurity spectral function $A(\omega)$
reconstructed in the HF and bare orbital bases at increasing subspace size,
against the exact (ED) result. (c) Pointwise error $|A-A_{\rm exact}|$ for
the two bases at matched ground pool size $128$. (d) Error
$\max_\omega|A-A_{\rm exact}|$ versus the fraction $d/D$ of the ground
sector retained; the dotted line is the convergence threshold, $0.05$.
(e) Augmentation cost: configurations added to the branch pool per
ground-pool configuration on applying the multi-mode impurity operator
$c_0=\sum_p w_p d_p$; the dotted line marks one child per parent.
(f) Error versus total shot count, with bands spanning the spread over
random seeds. (g) Error versus the pool size actually realized, comparing deterministic
(exact-amplitude) ranking, the oracle, with finite-shot sampling in each orbital basis. Threshold crossings give the pool size
needed for a target accuracy. (h) Cumulative Krylov weight of the ground sector versus configurations retained, ranked by
importance.}
\label{fig:L8U2}

\end{figure*}

\begin{figure*}
    \centering
    \begin{tikzpicture}
      \node[anchor=south west,inner sep=0] (img) at (0,0)
        {\includegraphics[width=0.75\linewidth]{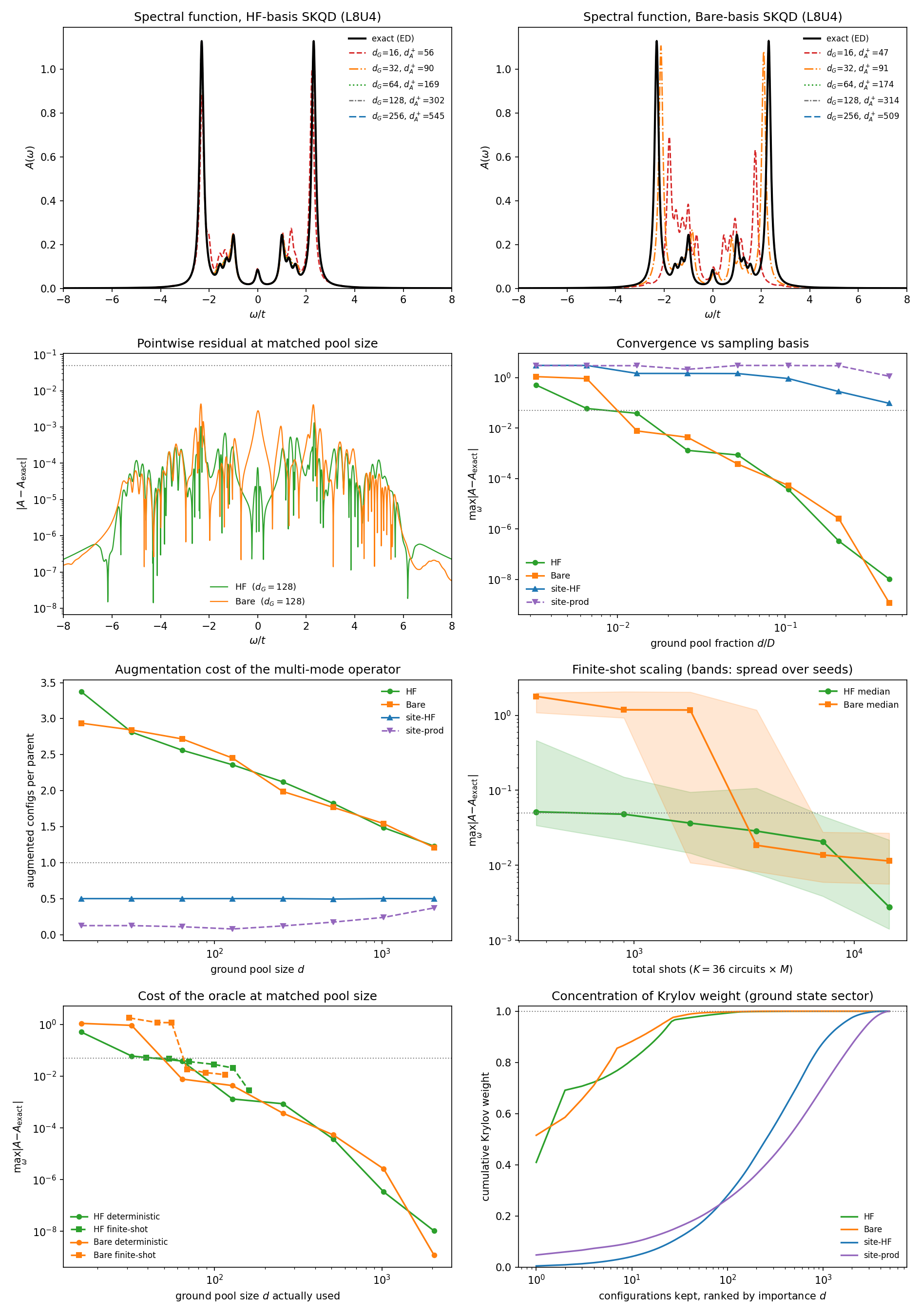}};
      \begin{scope}[x={(img.south east)},y={(img.north west)},
                    every node/.style={anchor=west,font=\small}]
        \node at (0.02,0.985) {(a)};  \node at (0.52,0.985) {(b)};
        \node at (0.02,0.740) {(c)};  \node at (0.52,0.740) {(d)};
        \node at (0.02,0.495) {(e)};  \node at (0.52,0.495) {(f)};
        \node at (0.02,0.250) {(g)};  \node at (0.52,0.250) {(h)};
      \end{scope}
    \end{tikzpicture}
   \caption{SKQD results for the particle--hole-symmetric SIAM chain at
$L=8$, $U=4$. (a),(b) Impurity spectral function $A(\omega)$
reconstructed in the HF and bare orbital bases at increasing subspace size,
against the exact (ED) result. (c) Pointwise error $|A-A_{\rm exact}|$ for
the two bases at matched ground pool size $128$. (d) Error
$\max_\omega|A-A_{\rm exact}|$ versus the fraction $d/D$ of the ground
sector retained; the dotted line is the convergence threshold, $0.05$.
(e) Augmentation cost: configurations added to the branch pool per
ground-pool configuration on applying the multi-mode impurity operator
$c_0=\sum_p w_p d_p$; the dotted line marks one child per parent.
(f) Error versus total shot count, with bands spanning the spread over
random seeds. (g) Error versus the pool size actually realized, comparing deterministic
(exact-amplitude) ranking, the oracle, with finite-shot sampling in each orbital basis. Threshold crossings give the pool size
needed for a target accuracy. (h) Cumulative Krylov weight of the ground sector versus configurations retained, ranked by
importance.}
\label{fig:L8U4}

\end{figure*}

\begin{figure*}
    \centering
    \begin{tikzpicture}
      \node[anchor=south west,inner sep=0] (img) at (0,0)
        {\includegraphics[width=0.75\linewidth]{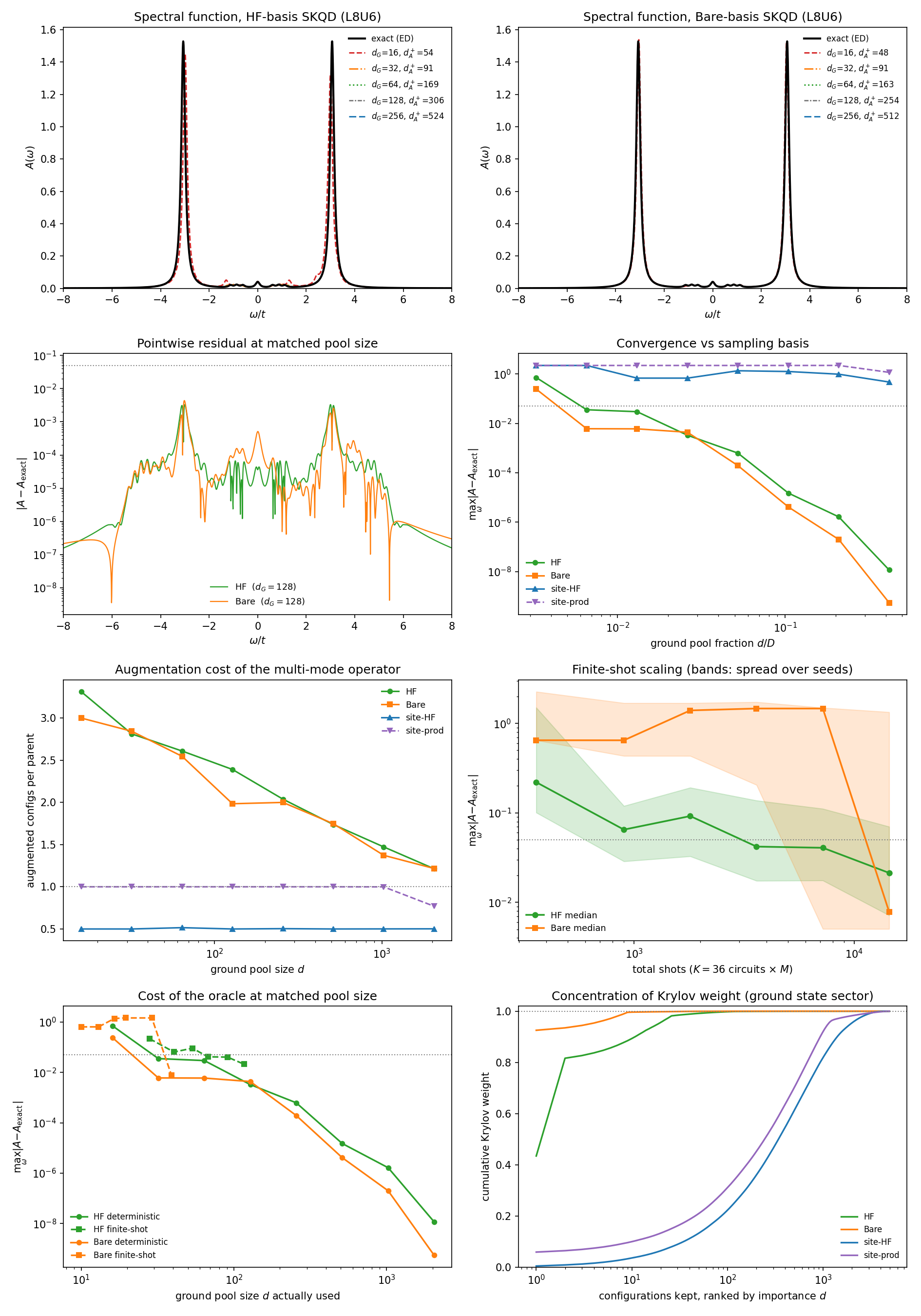}};
      \begin{scope}[x={(img.south east)},y={(img.north west)},
                    every node/.style={anchor=west,font=\small}]
        \node at (0.02,0.985) {(a)};  \node at (0.52,0.985) {(b)};
        \node at (0.02,0.740) {(c)};  \node at (0.52,0.740) {(d)};
        \node at (0.02,0.495) {(e)};  \node at (0.52,0.495) {(f)};
        \node at (0.02,0.250) {(g)};  \node at (0.52,0.250) {(h)};
      \end{scope}
    \end{tikzpicture}
   \caption{SKQD results for the particle--hole-symmetric SIAM chain at
$L=8$, $U=6$. (a),(b) Impurity spectral function $A(\omega)$
reconstructed in the HF and bare orbital bases at increasing subspace size,
against the exact (ED) result. (c) Pointwise error $|A-A_{\rm exact}|$ for
the two bases at matched ground pool size $128$. (d) Error
$\max_\omega|A-A_{\rm exact}|$ versus the fraction $d/D$ of the ground
sector retained; the dotted line is the convergence threshold, $0.05$.
(e) Augmentation cost: configurations added to the branch pool per
ground-pool configuration on applying the multi-mode impurity operator
$c_0=\sum_p w_p d_p$; the dotted line marks one child per parent.
(f) Error versus total shot count, with bands spanning the spread over
random seeds. (g) Error versus the pool size actually realized, comparing deterministic
(exact-amplitude) ranking, the oracle, with finite-shot sampling in each orbital basis. Threshold crossings give the pool size
needed for a target accuracy. (h) Cumulative Krylov weight of the ground sector versus configurations retained, ranked by
importance.}
\label{fig:L8U6}

\end{figure*}

Figures~\ref{fig:L10U2}--\ref{fig:L8U6} repeat the $L=12$ analysis of
Sec.~\ref{sec:Results} for chains of length $L=10$ and $L=8$, using
the bath parameters of Appendix~\ref{app:bath}. The figure layout,
reference states, and error measures are the same as in
Figs.~\ref{fig:L12U2}--\ref{fig:L12U6}; only the system size is changed.


\clearpage

\bibliography{references}
\clearpage
\appendix
\onecolumngrid
\end{document}